\documentclass[11pt]{article}
\usepackage{amssymb}
\usepackage{enumerate}
\usepackage{amsmath,amsthm}
\usepackage{natbib}
\usepackage{color}

\usepackage{ifpdf}
\ifpdf
\usepackage[pdftex]{graphicx}
\else
\usepackage[dvips]{graphicx}
\fi

\newcommand\E{\ensuremath{\mathbb{E}}}
\newcommand\R{\ensuremath{\mathbb{R}}}

\newcommand\F{\mathcal{F}}

\newcommand\la{\lambda}

\DeclareMathOperator*{\esssup}{ess\,sup}
\DeclareMathOperator*{\essinf}{ess\,inf}
\DeclareMathOperator*{\argmin}{arg\,min}

\newcommand\Ph{\ensuremath{\mathbb{P}}}
\newcommand\lah{\hat{\lambda}}
\newcommand\Wh{\hat{W}}

\def\Zh{Z^{\phi, \alpha}}

\def\L{{\mathcal L}}
\def\M{{\mathcal M}}
\def\R{{\mathbb R}} 
\def\P{{\mathbb P}} 
\newcommand{\EE}{{\mathord{I\kern -.33em E}}}
\def\E{{\EE}} 
\def\Fil{{\mathbb F}} 
\def\D{{\mathcal D}}

\def\B{{\mathcal B}}

\def\1{1{\hskip -3.3 pt}\hbox{I}}
\def\F{{\mathcal F}} %

\def\setT{{\mathcal T}}
\def\Qb{{\tilde{Q}}} %
\def\phib{{\tilde{\phi}}} %
\def\alpb{{\tilde{\alpha}}} %
\def\lab{{\tilde{\lambda}}} %
\def\hatP{\hat{P}} %
\def\byp{\tilde{P}} %
\def\mkp{P}
\def\Jam{{\mathcal J}}

\newtheorem{theorem}{Theorem}[section]

\newtheorem{corollary}[theorem]{Corollary}

\newtheorem{assumption}[theorem]{Assumption}

\newtheorem{proposition}[theorem]{Proposition}

\theoremstyle{remark}
\newtheorem{remark}[theorem]{Remark}
\newtheorem{example}[theorem]{Example}

\begin{document}

\title{Optimal Timing to Purchase Options}
\author{Tim Leung\thanks{Department of Applied Mathematics \& Statistics, Johns
    Hopkins University, Baltimore MD 21218; \mbox{{\em
        leung@jhu.edu}.} Work supported by NSF grant DMS-0908295 and Johns Hopkins Duncan Travel Fund.} \and Mike Ludkovski\thanks{Department of Statistics \& Applied Probability,
University of California Santa Barbara, Santa Barbara CA 93106; \mbox{{\em
        ludkovski@pstat.ucsb.edu}}.}}\date{\today} \maketitle

\begin{abstract}
We study the optimal timing of derivative purchases in incomplete markets.
In our model, an investor attempts to maximize the spread between her
model price and the offered market price through optimally timing her
purchase. Both the investor and the market value the options by
risk-neutral expectations but under different equivalent martingale
measures representing different market views. The structure of the
resulting optimal stopping problem depends on the interaction between the
respective market price of risk and the option payoff. In particular, a crucial
role is played by the delayed purchase premium that is related to the
stochastic bracket between the market price and the buyer's risk premia. Explicit characterization of
the purchase timing is given for two representative classes of Markovian
models: (i) defaultable equity models with local intensity; (ii) diffusion
stochastic volatility models. Several numerical examples are presented to
illustrate the results. Our model is also applicable to the optimal rolling of long-dated options and sequential buying and selling of options.
\end{abstract}

{\bf Keywords}: optimal stopping, delayed purchase premium, martingale measures, risk premia

\newpage

\section{Introduction}
In most financial markets, one fundamental problem for investors is to decide \emph{when} to buy a derivative at its current trading price. A potential buyer has the option to acquire the derivative immediately, or wait for a (possibly) better deal later. Naturally, the optimal timing for the derivative purchase involves comparing the buyer's subjective price and the prevailing trading price, which directly depend on the price of the underlying asset and the market views of the buyer and the market.

The majority of option pricing literature is concerned with sell-side perspective and focuses on hedging of options. In this view, the derivative contract is already given and the goal is to efficiently price it and then hedge it to monetize the transaction value with zero (or rather minimal) risk. In contrast, from a buy-side perspective (that of hedge fund managers, proprietary traders, etc.), the aim is to extract profit by finding \emph{mispriced} contracts in the market.
Portfolio managers will survey the entire traded derivative landscape to find options that from their view are improperly priced. They will then try to exploit this mismatch to make profit. Similarly, for over-the-counter (OTC) derivatives that are traded bilaterally off the exchange, the manager will look for a counterparty that offers an attractive price.

Consequently, two key aspects emerge. First, the market is naturally assumed to be \emph{incomplete}. Indeed, by the standard no-arbitrage pricing theory, the price of a derivative is given by the expected discounted payoff under some equivalent martingale measure, also known as the risk-neutral pricing measure. If the market is arbitrage-free and complete, there is only one pricing measure, and no ``mispricing'' is possible.
However, when the market is incomplete, there exist many candidate equivalent martingale measures that will yield no-arbitrage prices. Derivative buyers and sellers with different pricing measures (or market views) will assign different prices to derivatives over time. Therefore, the buyer's (resp.\ seller's) objective is to take advantage of the price discrepancy and optimally purchase (resp.\ sell) a contingent claim given the knowledge of the trading prices. Second, optimal \emph{timing} of trades is necessary to extract maximum profit. Indeed, even if a mispricing exists today, it is not clear whether it should be immediately exploited or rather one should wait for an even larger mispricing in the future. Thus, the time-dynamics of prices under different measures become crucial.

In this paper, we study the optimal purchasing decision from the
perspective of a derivative buyer. This leads to the analysis of a number of
optimal stopping problems over a finite horizon. As is common in the
literature, these problems do not admit closed-form solutions, so our focus
is to analyze the corresponding probabilistic representations and variational
inequalities, and illustrate the optimal purchasing strategies through
numerical examples. For instance, using a martingale argument, we can
deduce whether the buyer will purchase immediately or never purchase
based on the pricing measures used and the contract type (e.g.\ a Put or
Call), and determine what factors accelerate or delay the purchasing
decision. We also introduce the idea of the \emph{delayed purchase
premium} to provide alternative mathematical and financial explanations to
the buyer's purchase timing. We show that this delayed purchase premium
is closely related to the stochastic bracket between the market price and
the state price deflator, and it provides a connection between the families
of martingale measures and the properties of contract prices.

For buying American options, the buyer faces a two-stage optimal
stopping problem, in which the purchase date is first selected, followed by an option exercise date.
We find that the delayed purchase premium for an
American option  has a direct connection with its \emph{early exercise
premium} \citep{carr_jarrow_myneni92}.  In the
case of buying perpetual American Puts under a defaultable stock model,
we give explicit solutions for the option prices, the buyer's value function,
as well as the optimal purchase and exercise thresholds.

In incomplete markets, there are many candidate risk-neutral pricing measures that will yield no-arbitrage prices. Some well-known examples include the minimal martingale measure \citep{FollmerSchweizer1990}, the minimal entropy martingale measure \citep{FujiwaraMiyahara03,Frittelli00}, and the $q$-optimal martingale measure \citep{Hobson04}. In most practical models of incomplete markets, the various pricing measures are parameterized through the market price of risk of some nontraded factor. For the derivative purchase problem, it is then important to understand the dependence of prices on market price of risk, as well as the evolution of market price of risk over time. Two representative setups we will discuss below include (i) equity models subject to default risk and  (ii) stochastic volatility models with volatility driven by a second stochastic factor. In models of type (i), we will be concerned with market price of default risk; in models of type (ii) with market price of volatility risk.

To our knowledge, the purchase timing problem considered herein is new in the mathematical finance literature. As explained above, it links together the extensive body of research on representations of equivalent martingale measures in incomplete markets and the continuous-time optimal stopping models. We also draw upon results comparing option prices under different pricing measures, such as \cite{RomanoTouzi97,HHHS}. Some existing work similar in flavor to ours includes study of optimal static-dynamic hedges \citep{leungsircar2} and quasi-static hedging \citep{AllenPadovani02}. Finally, in a recent series of papers, \cite{PeskirSameePut} and \cite{PeskirSameeAsian} proposed a new financial engineering contract termed \emph{British options}. In those works the classical complete Black-Scholes market is considered and the payoff upon exercise can be viewed as the undiscounted price of the claim under the ``contract'' (and non-martingale) measure $P^{\mu}$.

The rest of the paper is organized as follows. In Section \ref{sec:model}, we setup our mathematical model and the main structural results. In Section \ref{sect-defaultS}, we consider optimal timing of purchases for derivatives written on defaultable stocks, while in Section \ref{sect-stochV} we consider buying options on stocks with stochastic volatility. Finally, Section \ref{sec:conclude} concludes and points out related problems where our analysis can also be applied.

\section{Problem Overview}\label{sec:model}
In the background, we fix an investment horizon with a finite terminal
time $T$, which is chosen to coincide with the expiration date of all
securities in our model. We assume a probability space $(\Omega, \F,
\Ph)$ equipped with a filtration $\Fil=(\F_t)_{0\leq t \leq T}$, which satisfies
the usual conditions of right continuity and completeness.

The financial market consists of one risky asset $S$ and the riskless money
market account with a constant interest rate $r\geq0$. For the purpose of
presenting the main ideas in this section, we work with a general
incomplete market, where the price process $S$ is a c\`{a}dl\`{a}g, locally
bounded $(\Ph, \Fil)$-semimartingale. Our detailed analysis of the problem
will be conducted under two specific market models, namely, (i) a
defaultable stock model where $S$ is a geometric Brownian motion up to
an exogenous default time (in Section \ref{sect-defaultS}), and (ii) a
diffusion stochastic volatility model (in Section \ref{sect-stochV}). We
assume that all market participants have access to the same information,
encoded in $\Fil$. It is possible, though beyond the scope of this paper, to
introduce price discrepancies due to incomplete information via filtration
enlargement/shrinkage (see e.g. \cite{GuoJarrow_incomp_info_2009} and
\cite{ElKaroui_J_J_aftdefault10}).

Let us consider a buyer of a European-style option written on underlying $S$ with some payoff function $F(\cdot)$ at expiration date $T$. As is standard in no-arbitrage pricing theory, the market price of a derivative is computed according to some \emph{pricing measure}, or \emph{equivalent martingale measure} (EMM), $Q$, which does not lead to arbitrage opportunities. Therefore, we consider the trading price process for the option $F$, denoted by $(P_t)_{0\leq t\leq T}$, as given by
\begin{equation}\label{pit}P_t= \E^{Q}\{e^{-r(T-t)}F(S_T)|\,\F_t\}, \qquad 0\leq t\leq T.\end{equation}

The buyer may view the market as a representative agent (the seller), who sells option $F$ at the ask price $P_t$ for $t\in [0,T]$. Depending on the setup, this option may or may not be liquidly traded. Unless the market is complete, there exists more than one no-arbitrage pricing measure. Hence, we assume that the buyer computes her own mark-to-model price $\byp_t$ of the option under another pricing measure $\Qb$, namely,
\begin{equation}\label{pt} \byp_t=\E^{{\Qb}}\{e^{-r(T-t)}F(S_T)|\,\F_t\}, \qquad 0\leq t\leq T.\end{equation}

In many parametric market models, the pricing measure is directly linked to the \emph{risk premia} assigned to the underlying sources of risk. This provides a natural explanation to the difference in pricing measures and derivative prices between the buyer and the market (or among market participants in general).

\subsection{The Buyer's Optimal Stopping Problem}\label{sec:model_Jt}
The buyer has the opportunity to purchase the European option at the market price $P_t$ at or before its expiration date. The set of admissible purchase times, denoted by $\setT$, consists of all stopping times with respect to $\Fil$ taking values in $[0,T]$. The buyer's objective is to determine the optimal stopping time $\tau$ that maximizes the \emph{spread} between her subjective price $\byp_\tau$ and the market price $P_\tau$. At time $t\in[0,T]$, she faces the optimal stopping problem:
\begin{equation}\label{Jt}J_t = \esssup_{\tau \in \setT_{t,T}} \E^{{\Qb}}\{e^{-r(\tau-t)} ({\byp_\tau}- {P_\tau}) |\,\F_t\},\end{equation}
where  $\setT_{t,T} \triangleq \{\tau \in \setT: t\leq \tau \leq T\}$.  The quantity $J_t$ is interpreted as the optimal spread between the model price $\byp$ and the market ask price $P$ and can be used for statistical arbitrage algorithms. Namely, various profit opportunities can be ranked according to their spreads $J_t$ since other things being equal, larger $J_t$ is more likely to generate trading profit (in a generic case where true model is unknown).

Since $T$ is itself a trivial stopping time and $P_T = \byp_T = F(S_T)$, it follows from \eqref{Jt} that $J_t\geq 0$ and $J_T=0$. Hence, $J_t$ can be viewed as an \emph{American spread option}.  Since at time $T$ the option expires and all market participants realize the same payoff, the choice $\tau=T$ means the buyer \emph{never} buys the option. For instance, when the market price is consistently higher than the buyer's price, i.e.,\,$P_t \geq \byp_t$ for $t\in [0,T]$, we have $\tau^*=T$ and $J_t \equiv 0$ (see also Remark \ref{remark:Vts}).

By substituting (\ref{pt}) into (\ref{Jt}), along with repeated conditioning, we simplify $J_t$ to
\begin{align}J_t & =\esssup_{\tau \in \setT_{t,T}} \E^{{\Qb}}\left\{e^{-r(\tau-t)}\E^{{\Qb}}\{e^{-r(T-\tau)}F(S_T)|\,\F_\tau\} - e^{-r(\tau-t)}P_\tau \,|\,\F_t\right\}
 =  \byp_t - V_t,\end{align} where
\begin{align} \label{valfn}V_t :=\essinf_{\tau \in \setT_{t,T}} \E^{{\Qb}}\left\{  e^{-r(\tau-t)}P_\tau |\,\F_t\right\}.\end{align}
Therefore, in order to determine the buyer's optimal purchase time for $J_t$, one can equivalently solve the \emph{cost minimization} problem represented by $V_t$ in (\ref{valfn}). In other words, the buyer selects the optimal purchase time that minimizes the expected discounted market price under her pricing measure $\Qb$. If there were no market for the option $F$, then the investor's cost would be $\byp_t$. By optimally purchasing from the market, the investor's reduced cost is $V_t$. Therefore, one can view $J_t= \byp_t - V_t$ as the benefit of market access to the buyer.

In addition, we observe the following \emph{put-call  parity}   in terms of
the optimal purchase strategies.
\begin{proposition}\label{prop-pc} The buyer's optimal strategies for buying a European Call and for buying a {European} Put, with the same underlying, strike and maturity, are identical. \end{proposition}
\begin{proof} Let $p_t$ and $c_t$, respectively, be the  market prices of a European Put and European Call on $S$ with the same strike and maturity. Applying the well-known model-free Put-Call parity $c_t  - p_t=  S_t - Ke^{-r(T-t)}$ and the fact that $(e^{-rt}S_t)_{t\geq 0}$ is a $(\Qb, \Fil)$-martingale, we obtain
\begin{align}
 \E^{{\Qb}}\left\{  e^{-r(\tau-t)}c_\tau |\,\F_t\right\} & = \E^{{\Qb}}\left\{  e^{-r(\tau-t)} (p_\tau + S_\tau - Ke^{-r(T-\tau)}) |\,\F_t\right\}\notag\\
 & = \E^{{\Qb}}\left\{  e^{-r(\tau-t)} p_\tau  |\,\F_t\right\} + S_t - Ke^{-r(T-t)}.
\end{align}Since the last two terms do not depend on the choice of $\tau$, it follows that  \[\argmin_{\tau\in \setT_{t,T}}\E^{{\Qb}}\left\{  e^{-r(\tau-t)}c_\tau |\,\F_t\right\} = \argmin_{\tau\in \setT_{t,T}}\E^{{\Qb}}\left\{  e^{-r(\tau-t)}p_\tau |\,\F_t\right\}.\]
\end{proof}

Our aim for the remainder of the paper is to characterize the optimal acquisition time $\tau^*$ corresponding to the optimal stopping problem in (\ref{valfn}) in terms of $\Qb$ and $Q$. An important question is under what conditions it is optimal to immediately buy the option from the market, or conversely never purchase it. Moreover,  we want to examine the market factors, in particular the option payoff shape, that influence the investor to buy the option earlier or later.

\subsection{$\tau$-Optimal Concatenation of Pricing Measures}
The {minimum cost} $V_t$ can be alternatively viewed as the risk-neutral price of the option $F$ under some special measure. To this end, we first denote the density processes associated with $Q$ and $\Qb$ (with respect to $\Ph$) by
 \begin{align}Z^m_t = \E\left\{\frac{dQ}{d\Ph}\,|\,\F_t\right\}, \quad \text{ and } \quad   Z^b_t = \E\left\{\frac{d\tilde{Q}}{d\Ph}\,|\,\F_t\right\}, \quad 0\leq t\leq T,
 \end{align}where the expectations are taken under the historical measure $\Ph$. Next, we consider a probability measure $Q^\tau$ that is identical to $\Qb$ up to the $\F$-stopping time $\tau$ and then coincides with $Q$ over $(\tau, T]$. Precisely, $Q^\tau$ is defined through its $\Ph$-density process, $\frac{d Q^\tau}{d\Ph} |_{\F_t} =: Z^\tau_t$, as
\begin{align}\label{Ztau}Z^\tau_t := Z^b_t\, \1_{[0,\tau)}(t) + Z^m_t\frac{Z^b_\tau }{Z^m_\tau}\,\1_{[\tau,T]}(t), \qquad 0\leq t\leq T. \end{align}
It is straightforward to check that $(Z^\tau_t)_{0\leq t\leq T}$ is a strictly positive $\Ph$-martingale and that $Q^\tau$ is again an EMM. The expression in \eqref{Ztau}  is referred to as the \emph{concatenation} of the density processes $Z^b$ and $Z^m$ (or equivalently, the concatenation of the EMMs $\Qb$ and $Q$);
see, for example, \cite{Delbaen-mstable06}, and \cite{Riedel_stopAmbig09}. We denote by $\M(Q, \Qb)= \{Q^\tau\}_{\tau\in\setT}$ the collection of EMMs generated by concatenating the EMMs $Q$ and $\Qb$, parameterized by a stopping time $\tau$.

\begin{proposition}\label{prop:concatenation} The minimum cost $V_t$ can be expressed as
\begin{align}V_t&=\essinf_{Q^\tau\in \M(Q, \Qb) } \E^{Q^\tau}\{  e^{-r(T-t)} F(S_T)\, |\,\F_t\}.\label{Vt2}\end{align}
\end{proposition}
\begin{proof}Applying (\ref{pit}) into (\ref{valfn}), we obtain
\begin{align} V_t & = \essinf_{\tau \in \setT_{t,T}} \E\Bigl\{  (Z^b_\tau/Z^b_t )\,e^{-r(\tau-t)} \E\{ (Z^m_T/Z^m_\tau ) e^{-r(T-\tau)}F(S_T)|\,\F_\tau\} |\,\F_t\Bigr\}\notag\\
& =\essinf_{\tau \in \setT_{t,T}} \E\left\{ \frac{Z^b_\tau}{Z^b_t } \frac{Z^m_T}{Z^m_\tau}\, e^{-r(T-t)} F(S_T)\, \big|\,\F_t\right\}\notag\\
&=\essinf_{Q^\tau\in \M(Q, \Qb) } \E^{Q^\tau}\{  e^{-r(T-t)} F(S_T)\, |\,\F_t\},\notag
\end{align}where the last equality follows from a change of measure using the fact that $Z^\tau_T = \frac{Z^b_\tau Z^m_T}{Z^m_\tau}$ and $Z^\tau_t = Z^b_t $.
\end{proof}

According to Proposition \ref{prop:concatenation}, the purchase timing flexibility allows the buyer to expand the space of pricing measures from her \emph{single} pricing measure $\Qb$ to the \emph{collection} of measures $\{Q^\tau\}_{\tau\in\setT}$ that is linked to the market measure $Q$ through concatenation. Note that all these candidate pricing measures coincide with the market measure $Q$ after time $\tau$. In particular, the choice of $\tau=t$ or $\tau =T$ corresponds to pricing under $Q$ or $\Qb$, respectively.  By choosing the purchase time, the buyer is in effect choosing the optimal time to adopt the market pricing measure. Related models of timing the adoption of market model risk in the context of irreversible investment have been considered in the real options literature; see \cite{AlvarezStenbacka04}.

\subsection{Delayed Purchase Premium}\label{sec:dpp}
From the optimal stopping problem in (\ref{Vt2}), we observe the inequality $V_t \leq P_t \wedge \byp_t$. This implies that the timing option necessarily reduces the buyer's valuation of the claim $F$ from $\byp_t$ to $V_t$ at any $t\leq T$.
In order to quantify this benefit of optimally waiting to purchase the option from the market, we define the buyer's \emph{delayed purchase premium} as \begin{align}L_t := P_t- V_t \ge 0, \label{DPP}\end{align} where $P_t$ is current cost of the option given in (see (\ref{pit})) and $V_t$ is the minimized cost (see (\ref{valfn})).

Recall that the optimal stopping time $\tau^*$ in \eqref{valfn} corresponds to the first time the value process equals the reward process \cite[Appendix D]{KaratzasShreve01}. Using \eqref{Vt2} and \eqref{DPP} we obtain
\begin{align}\label{tau-defn}
\tau^*_t &= \inf\{\,t\leq u\leq T \,:\, V_u = P_u\,\} \\
&=\inf\{\,t\leq u\leq T \,:\, L_u  = 0\,\} \notag.
\end{align}
As a result, the buyer will purchase the option from the market as soon as the delayed purchase premium diminishes to zero.

Let $Z_t = \E\{\frac{d \Qb}{d Q}|\F_t\}$ be the density process between the equivalent measures $\Qb$ and $Q$. We can re-express the minimal purchase cost $V_t$ as
\begin{align}
V_t = \essinf_{\tau\in \setT_{t,T}} \E^Q \left\{(Z_\tau/Z_t)\,e^{-r(\tau-t)}P_\tau \,|\,\F_t \right\}.\notag
\end{align}
Let $\hat{P}_t = e^{-r t}P_t$ be the discounted market price. Applying the integration-by-parts formula \citep[p.~83]{ProtterBook} to the semimartingale $Z \hat{P}$, we obtain
\begin{align}\label{ZP}
Z_\tau \hat{P}_\tau = Z_t \hat{P_t} + \int_t^\tau Z_{s-}\, d\hat{P}_s + \int_t^\tau \hat{P}_{s-}\, dZ_s + \int_t^\tau \,d[\hat{P}, Z]_s,
\end{align}
where $[\hat{P},Z]$ is the covariation process of $\hat{P}$ and $Z$. Since both $\hat{P}$ and $Z$ are $(Q, \Fil)$-local martingales and assuming enough integrability, this implies that
\begin{align}\label{bracket-dep}
L_t = P_t - V_t = \esssup_{\tau\in \setT_{t,T}} \E^Q \left\{ -(Z_t)^{-1} \int_t^\tau e^{-r (s-t)} d[P, Z]_s \,\big|\,\F_t \right\}.
\end{align}
 Thus, we see that the bracket $G_t := [P, Z]_t$, which we call the \emph{drift function}, plays a crucial role in determining the delayed purchase premium and, in view of \eqref{tau-defn}, the optimal purchase time. This observation will be key to our Theorems \ref{prop-exercise1} and \ref{prop-exer-vol1} below that explicitly derive and analyze $[P, Z]$ in specific Markovian models. Expression \eqref{bracket-dep} can also be interpreted as the covariation process between the buyer's state price deflator $e^{-r t}Z_t$ and the market price $P_t$.

\begin{remark}\label{rem:peskir}
The British options studied in \cite{PeskirSameePut,PeskirSameeAsian} have payoffs of the related form $P^{\mu}(t,S_t) := \E^{\mu} \{ F(S_T) \,|\,\F_t \}$, where the expectation is taken under a non-martingale ``contract'' probability measure $\Ph^{\mu}$. Working under the Black-Scholes model, Peskir and Samee also derive an expression similar to the drift function $G$ (see e.g.\ (3.18) in \cite{PeskirSameePut}). They also characterize the early-exercise premium representation and the corresponding exercise boundary via a nonlinear integral equation using the methods of \cite{Peskir-Shiryaev-book}.
\end{remark}

\subsection{Buying American Options}\label{sec:BuyAmer}
The optimal timing of derivative purchase can be extended to the case of
American options. Suppose the investor is considering to buy a
finite-maturity American option, with payoff $F(S_\tau)$ at any exercise
time $\tau \in \setT$. At time $t\in[0,T]$, the buyer's price and the market
price are respectively given by
\begin{align}\label{AmerM} \byp^A_t = \esssup_{\nu \in \setT_{t,T}} \E^\Qb \left\{ e^{-r(\nu-t)} F(S_\nu)\,|\,\F_t \right\},\quad \text{ and }\quad
P_t^A = \esssup_{\nu \in \setT_{t,T}} \E^Q \left\{ e^{-r(\nu-t)} F(S_\nu)\,|\,\F_t \right\}.
\end{align}
The buyer's objective is to maximize the spread between his own price and
the market quote:
\begin{align}J_t^A& =\esssup_{\tau \in \setT_{t,T}} \E^{{\Qb}}\left\{e^{-r(\tau-t)} (\byp^A_\tau - P^A_\tau ) \,|\,\F_t\right\}.\label{JAmer}
\end{align}

Since $T$ is itself a stopping time and $\byp^A_T = P^A_T=F(S_T)$, this
implies that $J_t^A \ge 0$. Hence, any candidate stopping time $\tau^-$
with $\Qb\{\byp^A_{\tau^-} \leq  P^A_{\tau^-} \cap \tau^- < T\}>0$ is suboptimal, being dominated
by  $\tau^+ := \tau^- \1_{\{\byp^A_{\tau^-} >  P^A_{\tau^-}\}} +
T \1_{\{\byp^A_{\tau^-} \leq  P^A_{\tau^-}\}}$. It follows that  $\byp^A_\tau
>P^A_\tau \geq F(S_\tau)$ at purchase time $\tau <T$, which means the buyer will hold on to
the American option for a positive amount of time after purchase, rather
than exercise it immediately. However, it is still possible
that $P^A_\tau = F(S_\tau)$, so that the buyer may purchase the option
even when the market price reflects a zero exercise premium.

In contrast to its European counterpart,  the optimal timing problem
(\ref{JAmer}) with American options is more difficult since it involves
optimal \emph{multiple} stopping, namely, the optimal purchase  followed
by the optimal exercise. Due to the optimal stopping problems nested inside
the expectation in (\ref{JAmer}), the simplification \eqref{valfn} does not
apply.

On the other hand,  the  American option price process $(\byp^A_t)_{t\geq
0}$ is a $(\Qb, \Fil)$-supermartingale and can be decomposed into the
European option price plus the early exercise premium:
\begin{align}\label{Amer_decomp}\byp^A_t = \byp_t + \tilde{\Lambda}_t.\end{align}
See, for example, \cite{Kramkov_decompose} in a general incomplete
market and \cite{ElKaroui1995} in models with  Brownian motions. Note
that $\byp$ is a $(\Qb, \Fil)$-martingale, and $\tilde{\Lambda}$ is a
positive $\Fil$-adapted decreasing process with $\tilde{\Lambda}_T=0$.

To measure the value of optimal timing to purchase an American option, we define the
delayed purchase premium  by\begin{align}\notag L^A_t & :=
J^A_t - (\byp^A_t-P^A_t) \\ \label{delay-Amer}  & = \esssup_{\tau\in \setT_{t,T}} \E^\Qb
\left\{e^{-r (\tau-t)} (\tilde{\Lambda}_\tau - {\Lambda}_\tau) - \int_t^\tau
e^{-r (s-t)}(Z_s)^{-1} d[P, Z]_s \,\big|\,\F_t \right\},
\end{align}where the second equality follows from \eqref{ZP} and \eqref{Amer_decomp}.
In contrast to the delayed purchase premium $L_t$ for European options in
\eqref{bracket-dep},  the delayed purchase premium $L^A_t$ depends on
the early exercise premia difference $\tilde{\Lambda} - {\Lambda}$ in
addition to the stochastic bracket $G := [P, Z]$ between the European
option market price $P$ and the density process $Z$. In terms of $L^A$,
the optimal purchase time is given by $\tau^{A*}_t = \inf\{t\leq u\leq T
\,:\, L^A_u = \tilde{\Lambda}_u - {\Lambda}_u\,\}.$ In particular, when
$\tilde{\Lambda}_t < {\Lambda}_t$ the option is not purchased since
$L^A_t \geq \max\{\tilde{\Lambda}_t - {\Lambda}_t, 0\}$ according to
\eqref{delay-Amer}.

  Under a defaultable equity model, we will provide an explicit solution to
  the problem of buying perpetual American puts, as well as analysis on the finite-maturity American
  puts in Section \ref{sect-amerperp}.

\section{Buying Options on a Defaultable Stock}\label{sect-defaultS} In this section, we study the option purchase problem under a one-factor reduced-form defaultable stock model. Under the historical measure $\Ph$, the defaultable stock price $S$ evolves according to
\begin{equation}dS_t = (\mu + \lah_t )S_t \, dt + \sigma S_t\, d\Wh_t - S_{t-}\,dN_t,\qquad S_0=s>0,
\end{equation}with constant drift $\mu$ and volatility $\sigma >0$. Here, $\Wh$ is a standard Brownian motion under $\Ph$, and $\lah$ is the intensity process for the single jump process $N$ under $\Ph$. Specifically, we define
$$
N_t =1_{\{ t \geq \tau^{\lah}\}}, \quad\text{and}\quad \tau^{\lah} = \inf \bigl\{ \,t : \int_0^t \lah_s \,ds > E \bigr\}, \qquad\text{where}\quad  E \sim \mathit{Exp}(1), \;E \perp \F^{\Wh},
$$
and $\lah$ is a positive $\F^S$-adapted process. At $\tau^{\lah}$, the stock price immediately drops to zero and remains there permanently, i.e.\ for a.e.~$\omega \in \Omega$, $S_{t}(\omega) = 0,  \forall t \ge \tau^{\lah}(\omega)$. We denote the filtration $\F_t = \F^S_t \vee \sigma(E)$ and assume it satisfies
the usual conditions of right continuity and completeness. The compensated $(\Ph,\Fil)$-martingale associated with $N$ is given by $\hat{M}_t = N_t -\int_0^t \lah_s\,ds$. Herein, we will consider Markovian local intensity of the form $\lah_t = \lah (t,S_t)$, for some bounded positive function $\lah(t,s)$. To summarize, $S$ follows a geometric Brownian motion until the default time $\tau^{\lah}$, with a local default intensity $\lah$. Similar equity models have been considered e.g.\ in \cite{Linetsky06}, and date back to \cite{Merton_jump1976}.

 To begin our analysis, we define the set of equivalent martingale measures (EMMs) and study the price dynamics of options written on $S$. Following the standard procedure in the literature (see, among others, \cite{JarrowLandoYu05} and \cite{BellamyJeanblanc00}), an EMM $Q^{\phi, \alpha}$ is defined through the Radon-Nikodym density $\frac{d Q^{\phi, \alpha}}{d \Ph}|_{\F_t} = \Zh_t$, where
\[\Zh_t= \mathcal{E}(-\phi\Wh)_t \, \mathcal{E} (\alpha \hat{M})_t,\] is a product of the Dol\'eans-Dade exponentials
\begin{align}\mathcal{E}(-\phi\Wh)_t &= \exp \left(-\frac{1}{2}\int_{0}^{t} \phi^2_{s} \,d s  -\int_{0}^{t}\phi_{s}d \Wh_s  \right), \qquad\text{and}\\
\mathcal{E} (\alpha \hat{M})_t &=  \exp\left(  \int_0^t \log \alpha_s \,dN_s -\int_0^t \lah(\alpha_s - 1) 1_{\{ s < \tau^{\lah}\} }\,ds \right).
\end{align}Here, $(\alpha_t)_{0\leq t\leq T}$ is a strictly positive bounded $\F_t$-predictable process which acts as a scaling factor for the default intensity, and $(\phi_t)_{0\leq t \leq T}$ is another bounded process found from the equation \begin{equation}\label{condphi}\phi_t = \frac{\mu - r - \lah_t (\alpha_t -1)}{\sigma}.\end{equation}

The process $\phi$ is commonly referred to as the market price of risk and $\alpha$ as the default risk premium. The condition (\ref{condphi}), which is common in jump-diffusion models (see \cite{BellamyJeanblanc00} and references therein),  ensures that the discounted stock price is a martingale under $Q^{\phi, \alpha}$. Indeed, by Girsanov Theorem, the evolution of $S$ under any EMM $Q^{\phi, \alpha}$ is given by
\begin{equation}dS_t = r S_t \, dt + \sigma S_t\, dW^{\phi,\alpha}_t - S_{t-}\,d M^{\phi,\alpha}_t,\qquad S_0=s>0,
\label{SDES}\end{equation}where $W^{\phi,\alpha}_t  =  \Wh_t+\int_0^t  \phi_u \,du$ is a $Q^{\phi, \alpha}$-Brownian motion, and $M^{\phi,\alpha}_t = N_t - \int_0^t \alpha_s\lah_s \,ds$ is a $Q^{\phi, \alpha}$-martingale. Therefore, the default intensity under $Q^{\phi,\alpha}$ is $\la^{\alpha}_t = \alpha_t \lah_t$ and
the discounted stock price $(e^{-rt}S_t)_{t\geq 0}$ is a $Q^{\phi, \alpha}$-martingale.

According to (\ref{condphi}), the set of the risk-neutral pricing measures can be viewed as being parameterized by the default risk premium $\alpha$ only. Herein, we will consider Markovian default risk premium of the form $\alpha_t = \alpha (t,S_t)$ for some bounded positive function $\alpha(t,s)$. This makes the entire model Markov with state space $E = [0,T] \times \R_+$ and
the risk-neutral price under any $Q^{\phi, \alpha}$ of an European option with terminal payoff $F(S_T)$ can be written as
\begin{align}
P(t,S_t) = \E^{Q^{\phi, \alpha}}\{ e^{-r(T-t)}F(S_T)\,|\, S_t\},\label{Price-def}
\end{align} where $P(t,s)$ is a deterministic function which depends on the choice of $\alpha$. The discounted option price $\hatP(t,S_t):=e^{-rt}P(t,S_t)$ is a $Q^{\phi,\alpha}$-martingale and satisfies the SDE
\begin{align}d \hatP(t,S_t) &= e^{-rt} \sigma S_t \frac{\partial P}{\partial s}(t,S_t)\, dW^Q_t + e^{-rt} (P(t,0) - P(t,S_{t-})) \, dM^Q_t\notag\\
&=  \sigma S_t \frac{\partial \hatP}{\partial s}(t,S_t)\, dW^Q_t +  (\hatP(t,0) - \hatP(t,S_{t-})) \, dN_t -  \la^\alpha(t,S_{t-}) \left(\hatP(t,0) - \hatP(t,S_{t-})\right)\,dt.
\end{align}
Moreover, by standard Feynman-Kac arguments, the option price function $P(t,s)$ solves the inhomogenous linear PDE problem:
\begin{align} \label{PDE_P}
\left\{ \begin{aligned}
 \frac{\partial P}{\partial t}(t,s) + \mathcal{L}_{\la^{ \alpha}} P(t,s) + \la^{\alpha}(t,s) P(t,0) &= 0,& \quad &(t,s)\in [0,T)\times (0, \infty),\\
P(t,0) &= e^{-r(T-t)} F(0),& \quad &t \in [0,T),\\
P(T,s) &= F(s),& \quad &s\in[0, \infty),\end{aligned}\right.
\end{align}where  $\la^{\alpha}(t,s) := \alpha(t,s) \lah(t,s)$ is the default intensity under $Q^{\phi, \alpha}$, and $\mathcal{L}_{\la^{\alpha}}$ is the second order differential operator defined by \begin{align}\mathcal{L}_{\la^{\alpha}} f := (r+\la^{\alpha}(t,s))s \frac{\partial f}{\partial s} + \frac{1}{2} \sigma^2 s^2 \frac{\partial^2 f}{\partial s^2} - (r+\la^{\alpha}(t,s))f.\label{Lf}\end{align}
The dynamics of $\hatP(t,S_t)$ will play a crucial role in the option buyer's optimal stopping problem which we discuss next. Let us point out that as long as there are no liquid  contracts for hedging the default time, such as credit default swaps, the option market remains incomplete. Thus, in this setup we can assume that all vanilla Calls/Puts are liquid, and their market prices can be used to calibrate the market measure $Q$.

\subsection{The Buyer's Optimal Purchase Timing}
Denote $Q = {Q}^{\phi,\alpha}$ and $\Qb = Q^{\phib,\alpb}$ to be the market and the buyer's pricing measures, respectively. The option prices under $Q$ and $\Qb$ are denoted by $P(t,s)$ and $\byp(t,s)$, and are different due to different default risk premia $\alpha$ and $\alpb$ assigned by the market and the buyer.

At time $t \leq T$, the buyer maximizes profit by solving the optimal stopping problem:
\begin{align}\label{Jt-def}J(t,s)&:= \sup_{\tau \in \setT_{t,T}} \E^{{\Qb}} \left\{e^{-r(\tau-t)} (\byp(\tau,S_\tau)- P(\tau,S_\tau)) \,|\,S_t=s \right\}\\
&= \sup_{\tau \in \setT_{t,T}} \E^{{\Qb}} \left\{e^{-r(\tau-t)} (\byp(\tau,S_\tau)- P(\tau,S_\tau))\,\1_{\{\tau < \tau^{\lab}\}} \,|\,S_t=s \right\},\notag\end{align}where $\tau^{\lab}$ is the default time of $S$ under $\Qb$.
The second equality follows from the fact that $S$ stays at zero past $\tau^{\lab}$ and $P(t,0) =\byp(t,0)= e^{-r(T-t)} F(0)$. When the stock defaults, we have $J(t,0)=0$ since all price discrepancies between the buyer and the market are eliminated. As a result, on the event $\{\tau \geq \tau^{\lab}\}$, the timing option has no value, and the buyer will not purchase the derivative. This is also consistent with practice because most derivatives stop trading after the underlying defaults.

By applying repeated conditioning to \eqref{Jt-def}, we obtain $J(t,s)  = \byp(t,s) - V(t,s)$, where
\begin{align}
V(t,s) &:=\inf_{ \tau \in \setT_{t,T}} \E^\Qb\{ e^{-r(\tau-t)}P(\tau,S_\tau)\,|\,S_t=s\}\label{def-V}\\
&=\inf_{ \tau \in \setT_{t,T}} \E^\Qb\left\{ e^{-r(\tau-t)}[P(\tau,S_\tau)\,\1_{\{\tau < \tau^{\lab}\}} +e^{-r(T-\tau)} F(0)\,\1_{\{\tau \geq \tau^{\lab}\}} ]\,|\,S_t=s\right\}.\notag
\end{align}

Note that $V(t,0)=P(t,0)=e^{-r(T-t)} F(0)$ in the case of default, so it
follows from \eqref{tau-defn} that $\tau^* = \inf\left\{ 0\leq t\leq T\,:\, V(t,
S_t) = P(t,S_t)\,\right\} \leq \tau^\lab$ a.s. The possibility of default implies
two scenarios: (i) in the event $\{\tau^* < \tau^\lab\}$, the buyer
purchases the option prior to default, and (ii) in the event of default, i.e.
$\{\tau^* = \tau^\lab\}$, the optimal timing problem is over and no
purchase takes place. The buyer's optimal timing is characterized by the
buy region $\B$ and the delay region $\D$, namely,
\begin{align}\label{buyregionV}\B &= \{(t,s)\in [0,T]\times (0,\infty)\,:\,  V(t,s)= P(t,s)\} \quad \text{and} \\
\D & =  \{(t,s)\in [0,T]\times  (0,\infty)\,:\,  V(t,s)  < P(t,s) \} = \{ (t,s)\in [0,T]\times  (0,\infty)\,:\, L(t,s) > 0 \}.
\end{align}

Furthermore, the variational inequality associated with $V(t,s)$ is
\begin{equation}
\min \Bigl( \frac{\partial V}{\partial t}(t,s) + \mathcal{L}_{\lab} \,V(t,s) + \lab(t,s) V(t,0), P(t,s) - V(t,s) \Bigr) = 0,
\label{VI_V}\end{equation}
for $(t,s) \in [0,T)\times \R_+$, with terminal condition $V(T,s) = F(s)$, for $s\in \R_+$.
Note that the market price $P(t,s)$ acts as the obstacle term in the variational inequality. Moreover, the default rates $\la(t,s)$ and $\lab(t,s)$ essentially act as state-dependent discount rates for the equations defining $P(t,s)$ and $V(t,s)$ respectively. Consequently, standard numerical tools for pricing of European/American-style options can be used to solve \eqref{PDE_P} and \eqref{VI_V}. Similar variational inequalities also arise in pricing American options under jump-diffusion models; see, for example, \cite{PhamAmer} and \cite{OksendalSulemBook}.

\begin{remark}\label{remark:Vts} If the market price always dominates the buyer's price, i.e., $P(t,s) \geq \byp(t,s)$ $\forall (t,s)$, then we can infer
from (\ref{Jt-def}) that $\tau^*=T$ and $J(t,s) =0$, which implies $V(t,s)
=\byp(t,s)$ (see (\ref{Jt})). We can also verify this by substituting $V(t,s)
=\byp(t,s)$ into the variational inequality (\ref{VI_V}) and using the PDE
(\ref{PDE_P}). For instance, this price dominance   can occur for American
Puts when the market default intensity dominates the buyer's, i.e.
$\la^\alpha(t,s) \geq\la^{\tilde{\alpha}}(t,s)$ $\forall (t,s)$; see
Proposition 5.1 of \cite{PhamAmer}.
\end{remark}

We now use \eqref{bracket-dep} to derive the drift function for the defaultable equity model and characterize the respective delayed purchase premium $L(t,s) = P(t,s)- V(t,s)$ (see (\ref{DPP})).

\begin{theorem}\label{prop-exercise1} Define the function\begin{align}\label{driftterm}G(t,s) := (\lab(t,s)-\la(t,s)) \bigr( s \frac{\partial  P}{\partial s} (t,s)+P(t,0)-P(t,s)\bigr).\end{align} If $G(t,s)\leq0$ for all $(t,s)\in [0,T]\times \R_+$, then it is optimal to never purchase the option, i.e. $\tau^*=T$ and $L(t,s) = P(t,s)-\byp(t,s) \ge 0$.  If $G(t,s) \geq 0$ for all $(t,s)\in [0,T]\times \R_+$, then it is optimal to purchase the option immediately, i.e. $\tau^*=t$ is optimal for $V(t,s)$, and $L(t,s) = 0$.
\end{theorem}
\begin{proof} Recall that $\phi_t$ and $\phib_t$ are the market prices of risk for the market and the buyer. It follows from the Girsanov Theorem that
\begin{align} dW^Q_t  &=  d\Wh_t+ \phi_t\, dt,\quad \text{and} \quad dW^\Qb_t =  d \Wh_t+ \phib_t \,dt,\label{WWdt}\notag
\end{align}where $W^Q$ is a $Q$-Brownian motion, and $W^\Qb$ is a $\Qb$-Brownian motion. This implies that \begin{align}dW^\Qb_t  = dW^Q_t + ( \phib_t-\phi_t) \,dt
& =  dW^Q_t + \frac{ \la_t-\lab_t}{\sigma}\, dt,\notag
 \end{align}
and the Radon-Nikodym derivative associated with the equivalent measures $Q$ and $\Qb$ is given by
\begin{align}
Z_t:=\frac{d\Qb}{dQ} \Big|_{\F_t}  = \mathcal{E}(- \varphi W^Q)_t \, \mathcal{E} (a M^Q)_t,
\end{align} where the processes $(\varphi_t)_{0\leq t\leq T}$ and $(a_t)_{0\leq t\leq T}$ are defined by $\varphi_t = \frac{\la_t-\lab_t}{\sigma}$ and $a_t =\frac{\lab_t}{\la_t}$. Since default risk premia are bounded, $Z$ is a true $Q$-martingale and
satisfies the SDE
\begin{align} dZ_t &= Z_{t-}\left[ -\varphi_t\,dW^Q_t +   (a_t -1) dM^Q_t\right]\notag\\
&=Z_{t-}\left[ -\varphi_t\,dW^Q_t +   (a_t -1) \,dN_t-(\lab_t -\la_t)\, dt \right],\end{align}
where $M^Q_t = N_t - \int_0^t \la_s\,ds$ is a $Q$-martingale.
Using Ito's formula, the dynamics of $Z \hatP$ under $Q$ are
\begin{align}
d(Z_t\hatP_t) & = \hatP_t \,dZ_t + Z_t\, d\hatP_t + d\hatP_t\, dZ_t\notag \\
& =\hatP_t \,dZ_t + Z_t\, d\hatP_t + Z_t (a_t-1) (\hatP(t,0) - \hatP(t,S_{t-}))\,dM^Q_t \notag \\
&\qquad  + Z_t (\lab_t-\la_t)  \bigr(  S_t \frac{\partial  \hatP}{\partial s} (t,S_t)+\hatP(t,0)- \hatP(t,S_{t-})\bigr)\, dt. \label{SDE_Y}
\end{align}Since $\hatP$, $Z$ and $M^Q$ are all $Q$-martingales, the drift of $d( Z_t\hatP_t)$ is the last $dt$ term. Therefore, the condition $G(t,s)\leq 0$ (resp. $G(t,s) \geq0$) implies that $Z \hatP$ is a $Q$-supermartingale (resp~$Q$-submartingale), and the result follows.

Finally, applying SDE (\ref{SDE_Y}) yields the buyer's {delayed purchase premium} as
\begin{align}L(t,s) &=  \sup_{\tau\in \setT_{t,T}} \E^\Qb\bigg\{- \int_{t}^{\tau}e^{-r(u-t)} G(u,S_u)\,du\,|\,S_t=s\bigg\},\label{FK-f}
\end{align}
which gives the conclusions of the Theorem in terms of $L(t,s)$.
\end{proof}

The drift function $G(t,s)$ is related to the \emph{gamma} or convexity of the option price $P(t,s)$. Indeed, if for each $t\in [0,T]$, $P(t,s)$ is convex in $s\in\R_+$, i.e.\, its gamma $P_{ss}(t,s) \geq 0$, then  \[\frac{\partial P}{\partial s}(t,s)  \ge \frac{P(t,s) - P(t,0)}{s}, \quad s\in \R_+,\] whereby the drift function takes the same sign as the difference in premiums, i.e. $G(t,s)(\lab(t,s)-\la(t,s))\geq 0$. Hence, the optimal purchase rule is simplified to a direct comparison of risk premia. In summary,

\begin{corollary}\label{cor:convex-def} Suppose the option price function $s\mapsto P(t,s)$ is convex for each $t\in [0,T]$. If $\lab(t,s)\leq \la(t,s)$  (resp.~$\lab(t,s) \geq \la(t,s)$) $\forall (t,s)\in [0,T]\times \R_+$, then $\tau^*=T$ (resp.~$\tau^*=0$).
\end{corollary}

As an application of Theorem \ref{prop-exercise1} and Corollary \ref{cor:convex-def}, we discuss an example with European Calls and Puts. Here, we assume $\la(t,s) = \la>0$. Then, the market Call and Put prices with strike $K$ are respectively given by \begin{align}\label{Ecall}C(t,s) &= C^{BS}(t,s; r+\la, \sigma, K, T), \\
P(t,s) &= P^{BS}(t,s; r+\la, \sigma, K, T)+ K e^{-r(T-t)}(1-e^{-\la(T-t)}),\label{Eput}\end{align} where $C^{BS}$ and $P^{BS}$ are the Black-Scholes pricing formulas for the Call and the Put. Both options are convex in $s$ and, applying Theorem \ref{prop-pc} and \eqref{driftterm} to \eqref{Ecall} and \eqref{Eput}, admit the same
drift function \begin{align}\label{Gts_callput}G(t,s) =  (\lab(t,s)-\la) K e^{-(r+\la)(T-t)} \Phi(d_2),\end{align}
where $\Phi$ is the standard Gaussian cdf
and $d_2$ is as in the classical Black-Scholes formula. By Corollary \ref{cor:convex-def}, if $\lab(t,s)\leq  \la$ for all $(t,s)$, then it is never optimal to purchase the Call or the Put, whereas if $\lab(t,s)\geq  \la$  for all $(t,s)$, then it is optimal to purchase them immediately.

 Theorem \ref{prop-exercise1} implies that to have a non-trivial purchase strategy, the expression $G(t,s)$ must change signs on $[0,T] \times \R_+$. For instance, if $s \frac{\partial  P}{\partial s} (t,s) + P(t,0) - P(t,s)\geq 0$, then the purchase strategy is trivial unless $\la(t,s) -\lab(t,s)$ can be both positive and negative. In other words, there must exist times and stock levels, such that the buyer's default intensity is less than the market's, and other times and stock levels such that the buyer's default intensity is larger than the market's. The location of the level set $\{ (t,s) : \la(t,s) = \lab(t,s) \}$ is then crucial for determining the optimal purchase boundary for the buyer.

The probabilistic representation  \eqref{FK-f} allows us to analyze the optimal purchase time $\tau^*$ via the premium $L(t,s)$.
Indeed, from \eqref{FK-f} it is clear that if $G(t,s) < 0$ then the buyer should postpone her purchase since positive infinitesimal ``rent'' can be derived by taking $\tau=t+\epsilon$ for $\epsilon$ sufficiently small in \eqref{FK-f}. Hence, for every $(t,s)$ in the buy region $\B$, we must have $G(t,s) \ge 0$. For instance, for a Call option we must have $\lab(t,s) - \la(t,s) \ge 0$ and the market must be underestimating the default intensity in the buy region. Furthermore, when the Call is
near expiry and $\lab(t,s) > \la(t,s)$, then $G(t,s) > 0$ and hence by continuity of $S$ until $\tau^{\lab}$, and $\lab$ being bounded, $L(t,s) = 0$ for $T-t$ small enough. Conversely, if $G(t,s) <0$, then $L(t,s) > 0$ near expiry and it follows that the critical  stock price $s^*(t)$ separating the buy and delay regions satisfies
$\lab(t,s^*(t)) = \la(t,s^*(t))$ in the limit $t \to T$.

Furthermore, if $G_1(t,s) \ge G_2(t,s)$ for all $(t,s)$, then the corresponding delayed purchase premia satisfy $L_1(t,s) < L_2(t,s)$. As a result, it is always optimal to purchase the derivative associated with $G_1$ \emph{before} that associated with $G_2$. We illustrate this observation through the following example.

 \begin{example}\label{bull-spread} (Call vs Bull Spread) Let us compare the buyer's optimal purchase timing between two bullish positions: a Call and a bull spread (also known as capped Call). First, we assume constant default intensities $\la$ and $\lab$ for the buyer and the market. The market price of the Call with strike $K$ is $C(t,s;K)$ as in (\ref{Ecall}), and its drift function $G(t,s;K)$ is given by (\ref{Gts_callput}). The market price of the bull spread with strikes $(K, K_h)$, $K<K_h$, is given by $B(t,s) :=C(t,s;K) - C(t,s;K_h)$. The corresponding drift function is $G_B(t,s) = G(t,s;K) - G(t,s; K_h)$, but it is not immediately clear from $G_B(t,s)$ what the buyer's optimal strategy is.

 Nevertheless, when $\lab\geq \la$, we have $G(t,s;K_h)\geq 0$ by (\ref{Gts_callput}), and therefore $G_B(t,s)\leq G(t,s;K)$. We can apply the observations above to conclude that the delayed purchase premium of the bull spread must dominate that of the Call, i.e.\ $L(t,s) \leq L_B(t,s)$. As a result, the optimal purchase time for the bull spread is always \emph{later} than the Call purchase time. In fact in this case, the buyer will buy the Call immediately, but may delay to buy the bull spread. By similar arguments, when $\lab \leq \la$, it follows that $G_B(t,s)\geq G(t,s;K)$, and the buyer will never purchase the Call but may buy the bull spread prior to expiration.  \end{example}

\subsection{Numerical Examples}\label{sect-numer}
In the cases where the purchase timing problem is nontrivial, we must
revert to numerical methods. Optimal stopping
problems on finite horizon generally do not admit closed-form solutions, but
have been extensively investigated in the literature. The defaultable equity
model above is one-dimensional in space and the most straightforward
algorithm is to solve the respective variational inequality. Note that we
have three possible formulations, namely solving for the profit spread
$J(t,s)$, the minimal purchase cost $V(t,s)$ or the delayed purchase
premium $L(t,s)$. The variational inequality for $V$ was given in \eqref{VI_V}
and  applying (\ref{PDE_P}) and (\ref{VI_V}) it follows that the variational inequality for $L$ is
\begin{equation}\begin{cases}\displaystyle
  \max \left( \frac{\partial  L}{\partial t}(t,s) + \mathcal{L}_\lab L(t,s) -G(t,s), -L(t,s) \right) = 0,\quad
\text{ for } (t,s) \in [0,T)\times \R_+,  \text{ and }\\
\displaystyle L(T,s) = 0, \text{  for } s\in \R_+.
\end{cases}\label{VI_f}\end{equation}
Both formulations yield the same exercise boundary. In the
examples below, we employed the standard implicit PSOR algorithm to
solve for $V(t,s)$ over a uniform grid (typically of size $10^3 \times
10^3$) on $[0,T] \times \R_+$ (see Ch.~9 of
\cite{WillmottHowisonDewynne95}). This method has the advantage that
a simple adjustment allows to compute $\byp(t,s)$ as well, and therefore derive
all the quantities of interest. Standard Dirichlet/von Neumann boundary
conditions were applied on the $S$-boundaries of the grid.

\begin{figure}[ht]
\begin{tabular*}{\textwidth}{lr}
\begin{minipage}{3.2in}
\center{\includegraphics[width=3.4in]{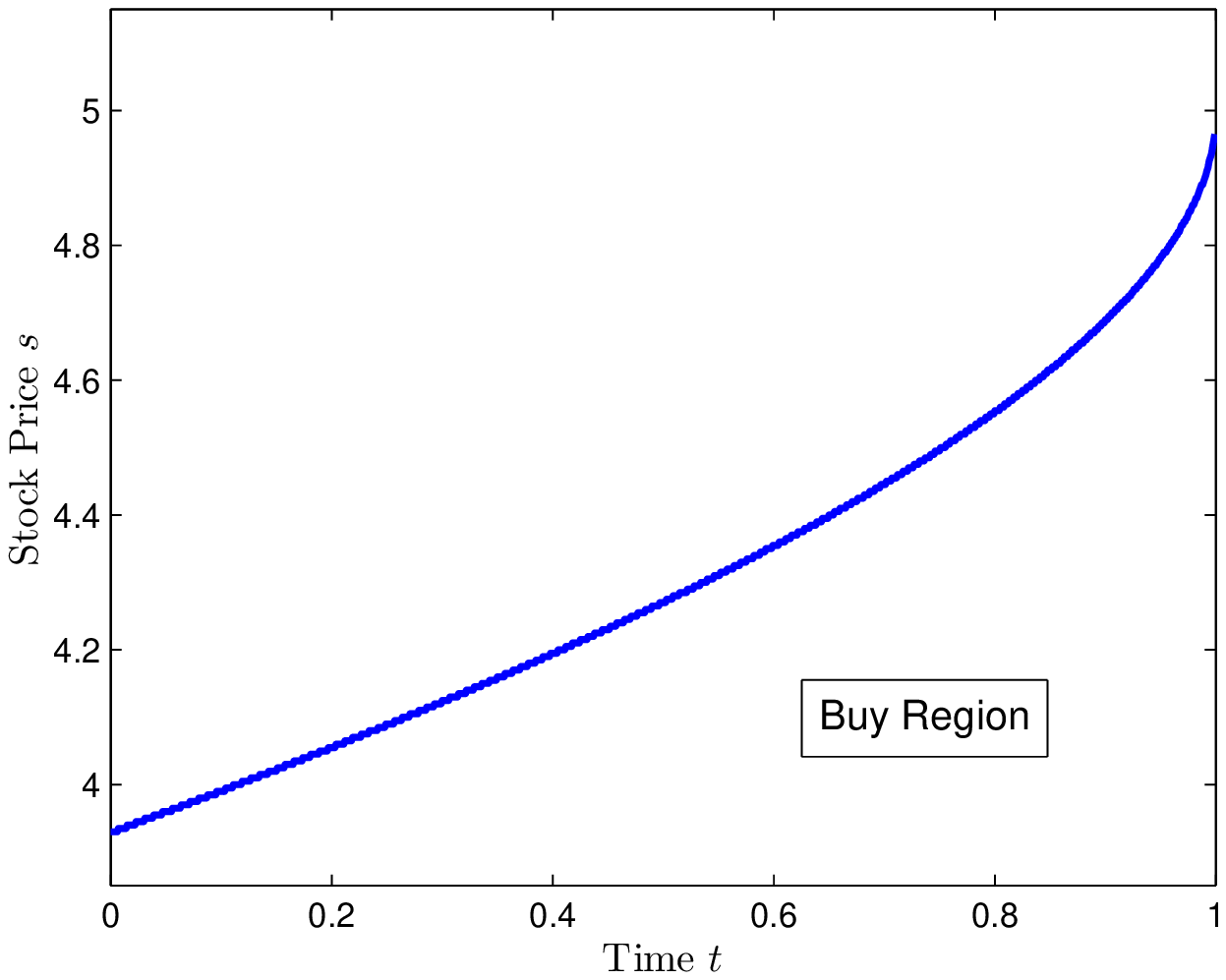}}
\end{minipage}
\begin{minipage}{3.2in}
\center{\includegraphics[width=3.4in]{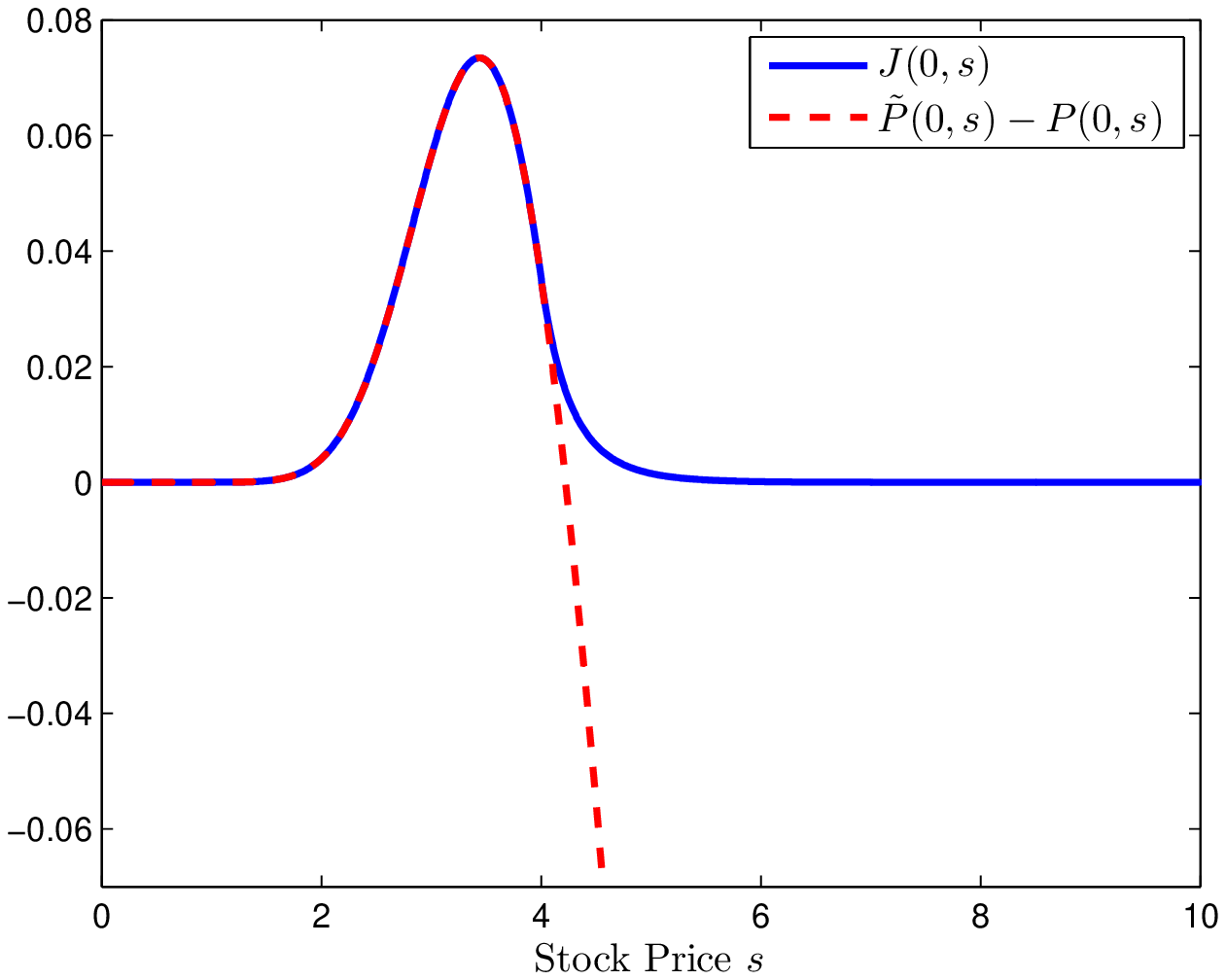}}
\end{minipage}
\end{tabular*}
\caption{\small{Call and Put purchase boundaries, with local intensity function $\la(t,s) = 0.2 $ and $\lab(t,s) = 0.2 e^{-0.2(s-K)}$. We take $r=0.05, \sigma=0.2, T=1$ and strike $K=5$.
\emph{Left panel}: purchase boundary for defaultable Call/Put.
\emph{Right panel}: profit spread for defaultable Put.
 \label{fig:put-call-power-default}}}
\end{figure}

Figure \ref{fig:put-call-power-default} illustrates the optimal purchase
boundary $t\mapsto s^*(t)$ that represents the critical stock value at
which the buyer should buy a European Put.  The buyer will buy as soon as
$S_t$ reaches $s^*(t)$ from above, but if default arrives first then $S$ will
jump across $s^*(t)$ to zero and no purchase will be made. In that
example, $\la(t,s) \equiv \la \ge \lab(t,s)$ for large $s$ and $\la(t,s)<
\lab(t,s)$ for small $s$. As explained above, by Put-Call parity, the purchase
boundary of the corresponding Call is the same. At maturity the purchase
boundary converges to $s^*(T) = K$, due to the fact that $\lab(K,s) =
\la(K,s)$.

Recall that the buyer's total profit is
\[J(t,s)=
\tilde{P}(t,s) - V(t,s) = [\tilde{P}(t,s) - P(t,s)] + L(t,s),\]
which decomposes into the current difference in valuations plus the delayed
purchase premium $L(t,s)$, see (\ref{DPP}). For instance with the
parameters of Figure \ref{fig:put-call-power-default} and initial stock price
$S_0 = 4.2$, the defaultable Put has market price $P(0,4.2) = 1.0542$
and investor's valuation of $\byp(0,4.2) = 1.0581$, so that a model-based
profit of $\$0.00393$ can be booked by buying this Put immediately. In
addition, we find that $L(0,4.2) = 0.0131$ so that another 1.3 cents (or
over $300\%$ of the above spread) can be gained by timing this purchase
optimally. The overall profit is therefore given by $J(0,4.2) =\$0.01704$.
Observe that the maximum profit of over 7 cents is realized around $S_0 =
3.5$ but in those cases it is optimal to lock it in immediately and $L(t,s) =
0$. The total gain from optimal timing of the derivative purchase can be
represented as
$$\min(P(t,s), \byp(t,s)) - V(t,s) = J(t,s) \wedge L(t,s) $$
and shows the profit obtained compared to the trivial strategies of $\tau^*
= 0$ and $\tau^* = T$.

\begin{figure}[ht]
\begin{tabular*}{\textwidth}{lr}
\begin{minipage}{3.2in}
\center{\includegraphics[width=3.4in]{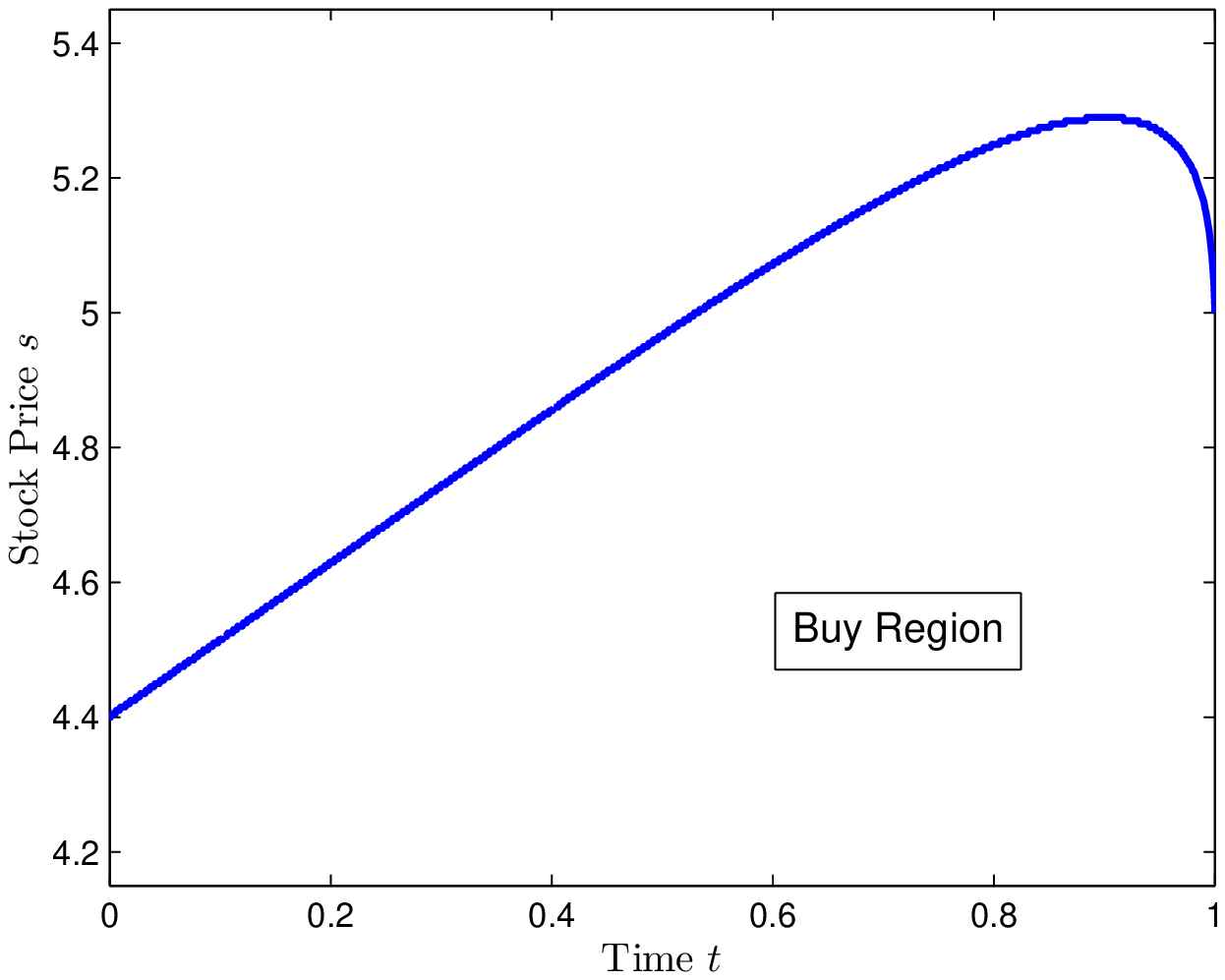}}
\end{minipage}
\begin{minipage}{3.2in}
\center{\includegraphics[width=3.4in]{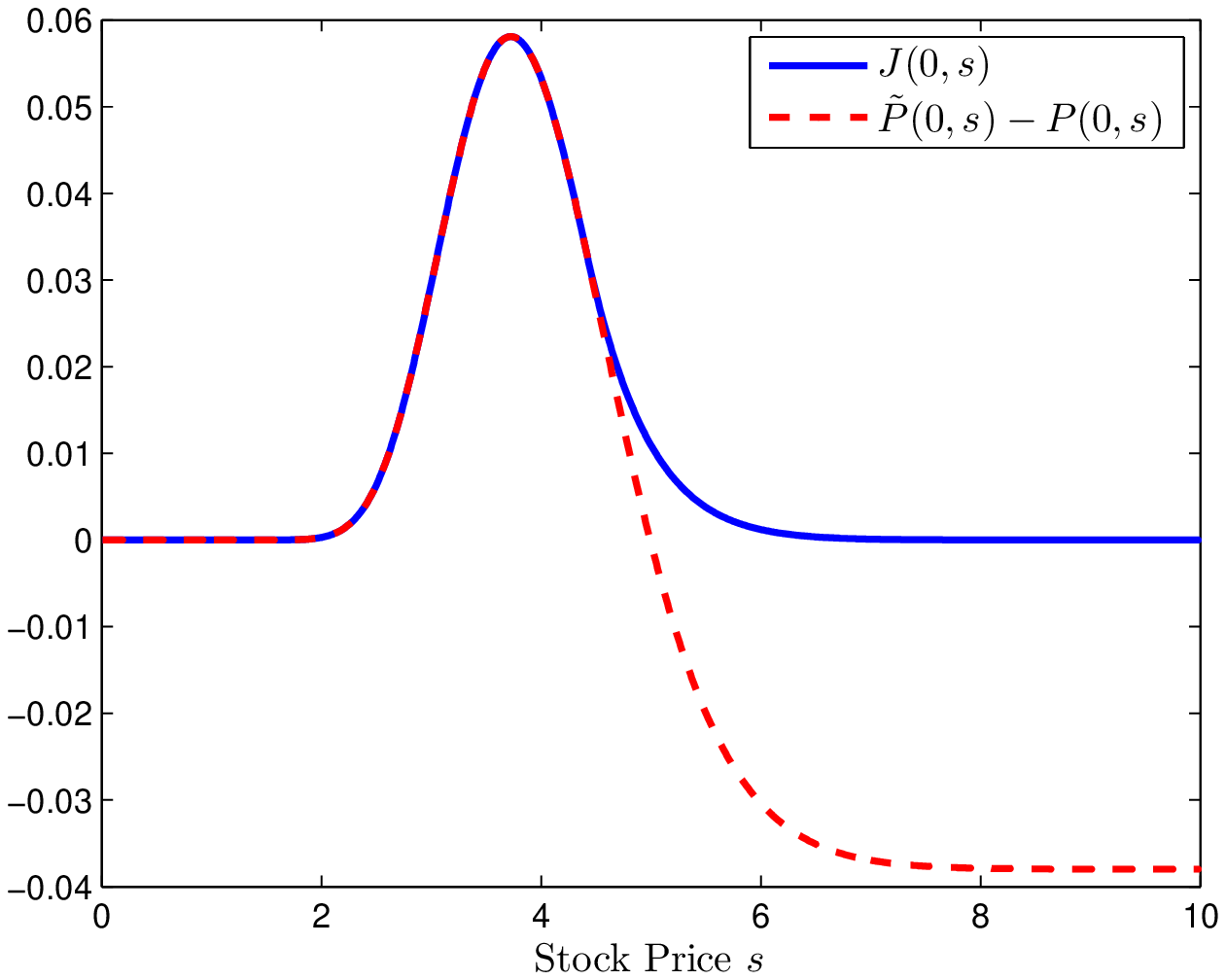}}
\end{minipage}
\end{tabular*}
\caption{\small{Digital Call purchase boundary with local default intensity functions $\la(t,s) = 0.2 $ and $\lab(t,s) = 0.25$. The remaining parameters are $r=0.05, \sigma=0.2, T=1$ and $K=5$. The digital Call pays out $F(s) = 1_{\{s>K\}}$. \emph{Left panel}: purchase boundary for defaultable digital Call.
\emph{Right panel}: corresponding value function $J(0,s)$ and price spread $\byp(0,s)-P(0,s)$.
\label{fig:digital-default}}}
\end{figure}

In Figure \ref{fig:digital-default}, we
consider a digital Call option with constant
default intensities $\la(t,s) \equiv \la$ and $\lab(t,s) \equiv \lab$ which
implies that the corresponding digital option prices are given by the classical
Black-Scholes formulas with discount rates $r+\la$ and $r+\lab$
respectively. The resulting drift function $G(t,s)$ is then
\begin{align*}
G(t,s) = (\lab - \la)e^{-(r+\la)(T-t)}\left(\phi(d_2) \frac{1}{\sigma \sqrt{T-t}} - \Phi(d_2)\right),
\end{align*}
where $\phi(\cdot)$ is the standard Gaussian density. $G(t,s)$ has
horizontal asymptotes $\lim_{s \to 0} G(t,s) = 0$ and $\lim_{s \to \infty}
G(t,s) = (\la - \lab)$ and, moreover, changes sign. As a result, the purchase
boundary $s^*(t)$, shown in Figure \ref{fig:digital-default} (left) is
non-trivial. Interestingly, this boundary is not monotone in $t$ and
moreover switches from being out-of-the-money for large $T-t$ to
in-the-money close to maturity. Similar non-monotonicity of $t \mapsto
s^*(t)$ is documented for British options, see Figure 5 in
\cite{PeskirSameePut}. The difference in prices $\byp(t,s) - P(t,s)$ also
exhibits a sign-change (right panel of Figure \ref{fig:digital-default}).

\begin{remark}
In all the examples above, both the purchase delay region $\D$ and the
buy region $\B$ were connected. This occurred because $G(t,s)$ was
monotone in $s$, which implies from \eqref{FK-f} that there is a simple
curve $t \mapsto s^*(t)$ separating $\B$ and $\D$. In more complicated
settings $G(t,s)$ may be changing signs several times which would lead to
multiple purchase boundaries and disconnected $\B$ and/or $\D$ regions.
\end{remark}

\subsection{Buying American Options}\label{sect-amerperp}
Continuing our discussion in Section \ref{sec:BuyAmer}, let us study the
optimal timing to buy American options under the defaultable equity model \eqref{SDES}.
To provide an example with closed-form solutions,  we first analyze the
purchase timing of a perpetual American Put with strike $K$. Assuming the market default intensity to be
a constant $\la$, standard calculations \cite[Ch.~2.7]{KaratzasShreve01} yield the market price and corresponding optimal exercise threshold
\begin{equation}\label{amerput_P}P(s) = \sup_{0\leq \tau <\infty} \E^Q\{ e^{-r\tau} (K-S_\tau)^+\} =
\begin{cases}\displaystyle
 \frac{ r K}{(r+\la)(\theta+1)}\left(\frac{s}{b^*}\right)^{-\theta}  + \frac{\la K}{r + \la}, \quad & \mbox{if } s > b^*, \\
\displaystyle K-s, & \mbox{if }s \leq b^*,
\end{cases}
\end{equation} where
\begin{align}
\label{amerput_s}b^* = \frac{2\,r }{2(r+\la)+ \sigma^2}K, \qquad \text{ and } \qquad  \theta = \frac{ 2(r+\la)}{\sigma^2}.\end{align}
Further assuming that the buyer also has a constant default intensity $\lab$ under $\Qb$, we may replace $\la$ with $\lab$ in \eqref{amerput_P}-\eqref{amerput_s} to obtain the buyer's value $\tilde{P}(s) = \sup_{0\leq \tau <\infty} \E^\Qb\{ e^{-r\tau} (K-S_\tau)^+\}$ and exercise threshold $\tilde{b}^*$.

\begin{figure}[ht]\center{\includegraphics[width=5in]{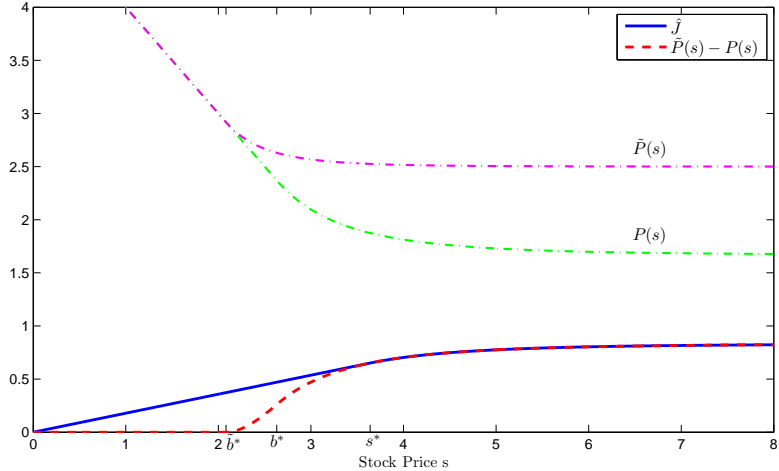}}
\caption{\small{Purchasing perpetual American Put. We take $\lab = 0.05$ and $\la=0.025$; the buyer's  perpetual American put price dominates the market price, so $\tilde{b}^*<b^*$. The price difference $\tilde{P}(s) - P(s)$ is increasing in $s$ (dashed curve at the bottom). The value function $\hat{J}(s)$ is the smallest concave majorant of $\tilde{P}(s) - P(s)$. Here, $\tilde{b}^*=2.0833$, ${b}^*=2.6316$, $s^*=3.6408$ and $\lim_{s \to \infty} \hat{J}(s) = 5/6$. Other parameters are $r=0.05, \sigma=0.2, K=5$.\label{fig-amerperp}}}
\end{figure}

The perpetual American option buyer's optimal stopping problem is
\[\hat{J}(s) =\sup_{0\leq \tau <\infty} \E^{\Qb} \left\{ e^{-r\tau } (\tilde{P}(S_\tau) - P(S_\tau))\right\}.\]
Note that $P(s)$ (resp. $\tilde{P}(s)$) is increasing and $b^*$ (resp.
$\tilde{b}^*$) is decreasing with respect to $\la$ (resp. $\lab$) (see Figure
\ref{fig-amerperp} (left)). If $\la \geq \lab$, then we have $\tilde{P}(s) -
P(s)\leq 0$ and $\hat{J}(s)=0$ for all $s\geq 0$. In this case, there is no
value in optimally timing the purchase. Henceforth, we will focus on  the
case with $\la <\lab$.

The payoff function $\tilde{P}(s) -
P(s)$ is increasing in $s$ but is neither convex nor concave. Nevertheless,
the buyer's optimal stopping problem admits a closed-form solution.

\begin{proposition}\label{prop-amerperp} Assume $\la <\lab$. The value of the timing option is
\begin{equation}\label{amer_J}\hat{J}(s) =
\begin{cases}\displaystyle
 A s,  \quad & \mbox{if } s < s^*, \\
\displaystyle \tilde{P}(s) - P(s) , \quad& \mbox{if }s \geq s^*,
\end{cases}
\end{equation}
with the constant $A$ given by \begin{align}\label{A}A = \frac{ r K \theta}{(r+\la)(\theta+1)s^*} \left(\frac{b^*}{s^*}\right)^{\theta}  -\frac{ r K \tilde{\theta}}{(r+\lab)(\tilde{\theta}+1)s^*} \left(\frac{\tilde{b}^*}{s^*}\right)^{\tilde{\theta}}, \end{align}
where $s^*$ is the optimal purchase threshold uniquely determined from the algebraic equation $B(s^*) = 0$ with
\begin{align}\label{sb}  B(s) := (r+\la)\left(\frac{\tilde{b}^*}{s}\right)^{\tilde{\theta}} - (r+\lab)\left(\frac{b^*}{s}\right)^{\theta}   + (\lab-\la).\end{align} Moreover, the thresholds are ordered by the inequality $\tilde{b}^* < b^* < s^*$.
\end{proposition}

\begin{proof} Let $\Jam(s)$ be the
conjectured solution in \eqref{amer_J}, which is simply the smallest
  concave majorant of the payoff function (see right panel of Figure
\ref{fig-amerperp}). To this end, the constants $s^*$ and $A$ are
chosen to satisfy the continuous-fit and smooth-fit conditions: $\Jam(s^*) =
A s^* = \tilde{P}(s^*) - P(s^*)$ and $\Jam^{'}(s^*)= A =\tilde{P}'(s^*) -
P'(s^*)$, which simplify to \eqref{A} and \eqref{sb}.  Furthermore,
$s^*$ exists and is unique and finite because the function $B(s)$ in \eqref{sb}
is strictly increasing for $s \geq b^*$, and satisfies $B(b^*) <0$ and $\lim_{s \to \infty} B(s) = \lab -\la > 0$.

By direct substitution and computation, we verify that $\Jam(s)$ satisfies   the variational inequality \begin{align} \label{VI_amerperp}\max \left(\frac{\sigma^2 s^2}{2} \Jam^{''}(s) +  (r + \lab) s \Jam^{'}(s) - (r+\lab) \Jam(s), \tilde{P}(s)-P(s)-\Jam(s) \right) =  0,
\end{align} for $s > 0$, with boundary condition $\Jam(0)=0$.
This implies that $(e^{-rt} \Jam(S_t))_{t \geq 0}$ is a (bounded) $(\Qb,
\Fil)$-supermartingale. Hence,   for any stopping time $\tau$,
 \begin{align}  \Jam(s) \geq \E^{\Qb} \{ e^{-r\tau } \Jam(S_\tau)\} \geq \E^{\Qb} \{ e^{-r\tau } (\tilde{P}(S_\tau) - P(S_\tau))\}. \label{amer_ineq}\end{align} Maximizing over $\tau$, we get $\Jam(s) \geq \hat{J}(s)$.  On the other hand, \eqref{amer_ineq} is an equality for the admissible stopping time $\tau = \inf\{ t\geq 0 : S_t \leq s^*\}$, which yields $\Jam(s) \leq \hat{J}(s)$.
\end{proof}

 Proposition \ref{prop-amerperp} is illustrated in Figure \ref{fig-amerperp}. We observe  that the value $\hat{J}(s)$ is linear in $s$ in the continuation region $[0,s^*]$ and increasing concave in $s$ in the exercise region $(s^*,\infty)$. It also admits the constant upper bound $\lim_{s\to \infty}\hat{J}(s) =(\frac{\lab}{r+\lab} - \frac{\la}{r+\la})K >0$. If the initial stock price $s\in(0,s^*)$, then the buyer will wait till the stock price $S$ hits the upper level $s^*$ to buy the perpetual American put before exercising it at a lower level $\tilde{b}^*$.

For perpetual American Calls written on  $S$ in (\ref{SDES}), the timing
problem is not well-defined. Indeed, the discounted price process
 $(e^{-rt}S_t)_{t\geq 0}$ is a martingale under $Q$ and $\Qb$, so it follows
 from Jensen's inequality that $(e^{-rt} (S_t-K)^+)_{t\geq 0}$ is a
 submartingale under $Q$ and $\Qb$. Therefore, the optimal policy for either the market or the buyer is
 to hold on to the Call forever.

Next, we turn our attention to the case of buying a finite-maturity
American Put. Denote by  $P^A(t,s)$ and $\byp^A(t,s)$, respectively, the market price and the buyer's price for the same American Put with payoff $F(s) = (K-s)^+$. The classical early exercise premium decomposition gives that
\begin{align} P^A(t, s) = P(t,s) + \Lambda(t,s)
&= P(t,s)+rK \int_t^T e^{-r(u-t)} Q_{t,s}\{  P^A(u, S_u) = F(S_u) \}\,du,\end{align}
and similarly for $\tilde{P}^A(t,s)$. The representation \eqref{delay-Amer} then implies that
\begin{multline}
L^A(t,s) =  \sup_{\tau\in \setT_{t,T}} \E^\Qb\bigg\{  - \int_{t}^{\tau}e^{-r(u-t)} G(u,S_u)\,du + \\ rK \int_\tau^T e^{-r(u-t)} \left[\Qb_{\tau, S_\tau}\{  \byp^A(u, S_u) = F(S_u) \} - Q_{\tau, S_\tau}\{  P^A(u, S_u) = F(S_u) \}\right]\,du\,|\,S_t=s\bigg\}.\label{Lts-amer2}
\end{multline}

When $\la$ and $\lab$ are constant, then the optimal strategy is to exercise the underlying American Put as soon as the stock reaches (from above) the exercise boundary $b^*(t)$, $t\in [0,T]$ (see Proposition 2.2 of \cite{PhamAmer}). As a result, $Q_{t,s}\{  P^A(u, S_u) = F(S_u) \} =Q_{t,s}\{S_u \leq  b^*(u) \}$ and the delayed purchase premium depends on the different probabilities under $Q$ and $\Qb$ that $S$ stays in the respective exercise regions. Note that the stock price may spend some time in the exercise regions before the buyer decides to purchase the option.

\begin{figure}[h]
\begin{tabular*}{\textwidth}{lr}
\begin{minipage}{3.2in}
\center{\includegraphics[width=3.4in]{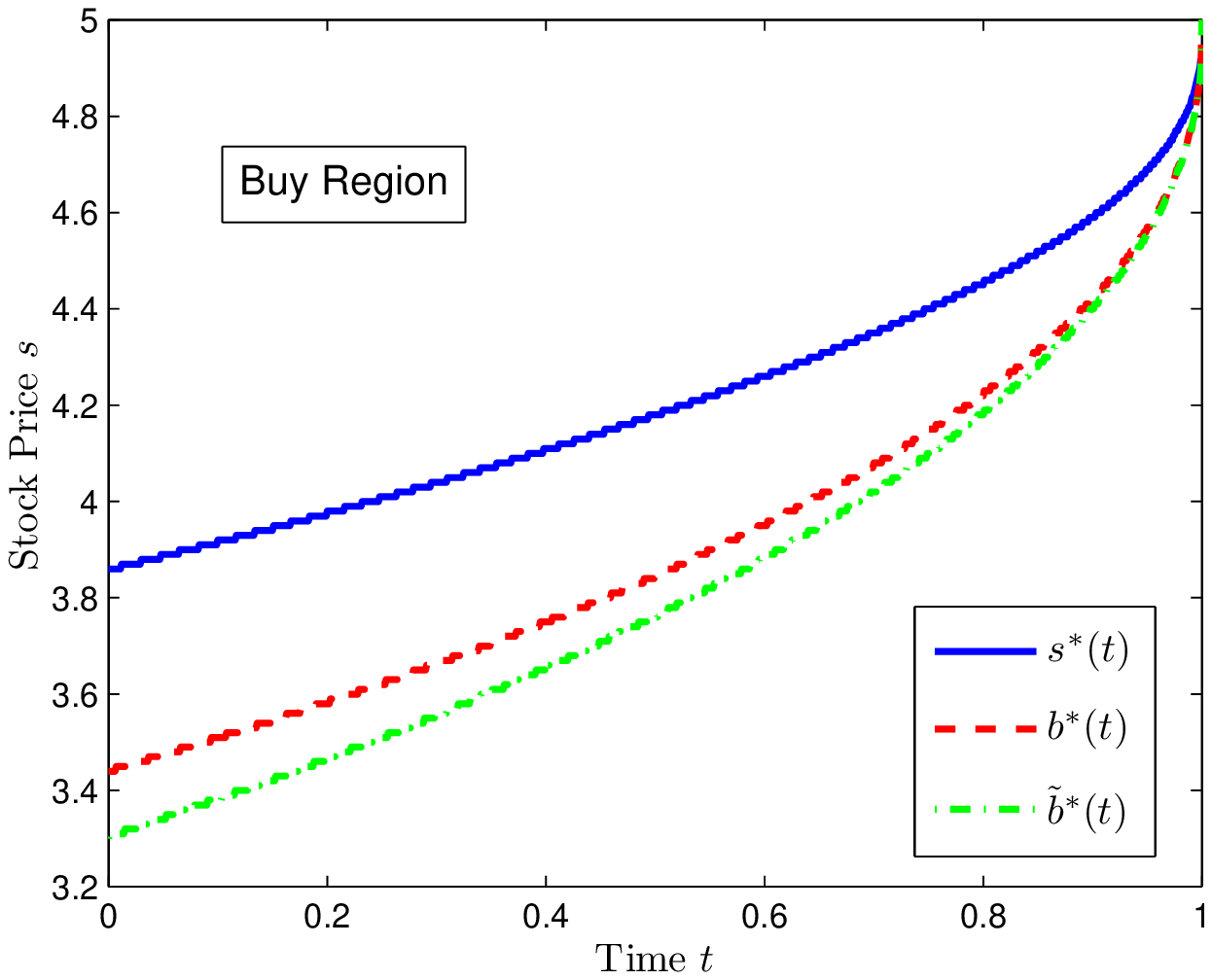}}
\end{minipage}
\begin{minipage}{3.2in}
\center{\includegraphics[width=3.4in]{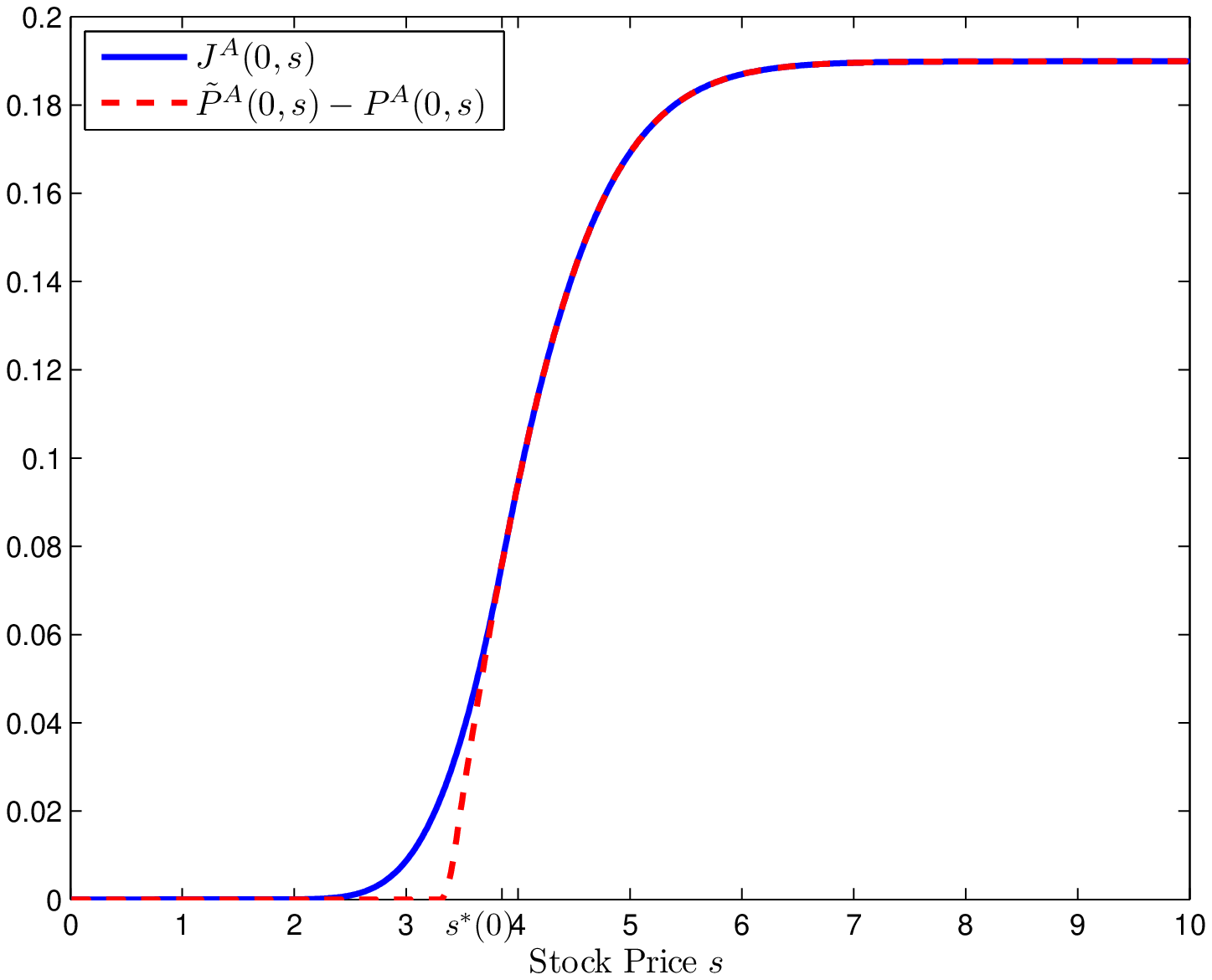}}
\end{minipage}
\end{tabular*}
\caption{\small{American Put purchase and exercise boundaries, with local default intensity functions $\la(t,s) = 0.2 $ and $\lab(t,s) = 0.25$. The other parameters are $r=0.05, \sigma=0.2, T=1$ and $K=5$. \emph{Left panel}: Solid line shows the purchase boundary $s^*(t)$; dashed line shows the market exercise boundary $b^*(t)$ and the dash-dotted line shows the investor's exercise boundary $\tilde{b}^*(t)$. \emph{Right panel}: the value function $J^A(0,s)$ and the price spread $\byp^A(0,s)-P^A(0,s)$.
\label{fig:amput}}}
\end{figure}

In Figure \ref{fig:amput}, we illustrate the optimal purchase
boundary. In this example, the default intensities are constant, and so the
underlying optimal exercise problems for $P^A$ and $\byp^A$ are
essentially identical to the classical American Put under the Black-Scholes
model with discount rate $r+\la$ (resp.~ $r+\lab$). The corresponding Put
{exercise} boundaries, denoted $b^*(t)$ and $\tilde{b}^*(t)$, are
shown in Figure \ref{fig:amput}. Since $\lab > \la$, we have $\byp^A(t,s)\ge
P^A(t,s)$ and $b^*(t) > \tilde{b}^*(t)$ for all $t,s$. The buy region $\B$ in
Figure \ref{fig:amput} is  {above} the \emph{purchase} boundary denoted
by $s^*(t)$, so that American Put is purchased on an up-tick.  Intuitively,
deep in-the-money the Put should be exercised under both EMMs, so that
$\byp^A(t,s) = P^A(t,s)$ and no profit spread is available. Conversely,
out-of-the-money $\byp^A(t,s)-P^A(t,s)$ is positive and concave in $s$ (Figure
\ref{fig:amput} right) and in the spirit of Corollary \ref{cor:convex-def} it is
optimal to purchase the American Put immediately. As a result, $s^*(t)$
lies slightly in-the-money, and for $S_0 \in (b^*(0), s^*(0))$ it is possible
that $\tau^*
> \tilde{\nu}^*_0 \vee \nu^*_0$, i.e.\ the Put is purchased after its
original exercise date under either EMM.

In this example, since $\lab > \la$, the European Put would be purchased immediately and $L(t,s)=0\, \forall (t,s)$ by Corollary \ref{cor:convex-def}. In contrast, the American Put's delayed purchase premium $L^A(t,s)$ is positive when $s < s^*(t)$ (Figure \ref{fig:amput} right), the purchase is delayed and the profit spread is larger $J^A(t,s) \ge J(t,s)$.

\section{Buying Options Under Stochastic Volatility}\label{sect-stochV} In this section, we study the problem of optimally buying an option under stochastic volatility. Under the historical measure $\Ph$ we consider a Markovian stochastic volatility model, where the underlying stock price $S$ and stochastic process $Y$ solve the SDEs:\begin{align}
dS_t &= S_t \, (\mu(t,Y_t) \,dt + \sigma(Y_t) \,dW_t),\label{SDESvol}\\
dY_t &= b(t,Y_t)\, dt + c(t,Y_t) \,(\rho dW_t+\hat{\rho} d\hat{W}_t).\label{SDEY}
\end{align} In \eqref{SDESvol}-\eqref{SDEY}, $W$ and $\hat{W}$ are two independent standard Brownian motions defined on $(\Omega, \F,(\F_t), \P)$, where $\F_t$ is taken to be the augmented $\sigma$-algebra generated by $((W_u, \hat{W}_u) ;0\leq u\leq t )$. The growth rate $\mu(t,Y_t)$ and the positive volatility coefficient $\sigma(Y_t)$ are driven by the \emph{non-traded} stochastic factor $Y$. We model the correlation between $S$ and $Y$ through the coefficient $\rho \in (-1,1)$ and set $\hat{\rho}= \sqrt{1-\rho^2}$.

\begin{assumption}\label{assumevol}(1) The volatility function $\sigma(\cdot)$ is Lipschitz, $C^1$-differentiable, and bounded above and below away from zero. (2) the functions $\mu(\cdot, \cdot)$ and $c(\cdot, \cdot)$ are bounded Lipschitz on $[0,T]\times\R$, with $c(\cdot,\cdot)\geq0$. (3) $b(\cdot, \cdot)$ is Lipschitz on $[0,T]\times\R$.
\end{assumption}
The stochastic volatility model in \eqref{SDESvol}-\eqref{SDEY} as well as Assumption \ref{assumevol} are adopted from the more general setups in \cite{RomanoTouzi97} and \cite{Sircarbook}.

Let $(\phi_t)_{0\leq t\leq T}$ be a bounded $\F$-progressively measurable process. Then, we can define an equivalent martingale measure $Q^\phi$ by \begin{equation}\label{Zlocalmtg}\frac{d Q^\phi }{d\P} \Big|_{\F_t}= \exp\left(- \frac{1}{2}\int_{0}^{t}
  \left(\kappa(s,Y_s)^2 +  \phi_s^2\right) \,ds -   \int_{0}^{t}
  \kappa(s,Y_s)\,d{W}_s - \int_{0}^{t}
  \phi_s\,d\hat{W}_s\right),\end{equation} where $\kappa(t,y) =\frac{\mu(t,y)-r}{\sigma(y)}$ is the bounded Sharpe ratio of $S$. By Girsanov's change of measure, the dynamics of $S$ and $Y$ under $Q^\phi$ are given by
\begin{align}
dS_t &= S_t \,  (\,r\,dt + \sigma(Y_t)\,dW^\phi_t),\label{SDESphi}\\
dY_t &= \left[\,b(t,Y_t)- \rho c(t,Y_t) \kappa(t,Y_t)  - \hat{\rho}c(t,Y_t) \phi_t   \, \right] \, dt + c(t,Y_t) \,(\rho dW^\phi_t+\hat{\rho} d\hat{W}^\phi_t),\label{SDEYphi}
\end{align}
where
\begin{align}\label{WWphi}W^{\phi}_t  = W_t + \int_0^t \kappa(s,Y_s)ds,\quad \text{ and }\quad
\hat{W}^\phi_t = \hat{W}_t + \int_{0}^{t}\phi_s\,ds\end{align} are independent
$Q^\phi$-Brownian motions. Therefore, the process $\phi$ parametrizes the set of pricing measures, and is typically called the\emph{ volatility risk premium}. In particular, when the risk premium is $\phi=0$, the associated measure $Q^0$ is the well-known minimal martingale measure (MMM) (see \cite{FollmerSchweizer1990}). The intuitive effect of $\phi$ is to modify the drift of $Y$ as observed in (\ref{SDEYphi}). Therefore, for options with positive dependence on volatility (such as those with convex payoffs and $\sigma'(\cdot)>0$, see \cite{RomanoTouzi97}), larger risk premium $\phi$ reduces the drift of $Y$ and hence is expected to decrease the option price. To this end, \cite{HHHS} have analyzed the ordering of option prices by risk premium under a stochastic volatility model (see also \cite{HenderHobs_JumpCompare} for jump-diffusions). The price ordering will also play a role in the buyer's optimal purchase decision.

\subsection{The Buyer's Optimal Purchase Timing}
For our analysis, we consider Markovian risk premia for the buyer and the market. Specifically, we let $\phib_t = \phib(t,S_t,Y_t)$ and $\phi_t = \phi(t,S_t,Y_t)$, for bounded continuous functions $\phib(t,s,y)$ and $\phi(t,s,y)$, which correspond to the buyer's measures $\Qb$ and the market measure $Q$ respectively. The option in question has a payoff $F(S_T)$ at expiration date $T$. The nontradability of $Y$ makes it impossible to completely replicate the option payoff by trading in $S$ and the money market account, so the market is incomplete. The
buyer's price and the market price are computed under their respective measures, namely, \begin{align}\byp(t,s,y) &= \E^{\Qb} \left\{e^{-r(T-t)}F(S_T)\,|S_t=s, Y_t=y \right\},\\
\text{and}\quad \mkp(t,s,y) &= \E^{Q} \left\{e^{-r(T-t)}F(S_T)\,|S_t=s, Y_t=y
\right\}.\label{price_vol}\end{align}
The buyer's objective is to solve the optimal stopping problem
\begin{align}V(t,s,y) = \inf_{\tau\in \setT_{t,T}}\E^{\Qb} \left\{ e^{-r(\tau-t)}P(\tau, S_\tau, Y_\tau)|S_t =s, Y_t=y \right\}.\label{Vvol}\end{align}
With this, the buyer's delayed purchase premium is given by $L(t,s,y) = P(t,s,y) - V(t,s,y)$. The buyer's optimal timing naturally depends on the option's market price $P(t,s,y)$ as well as the risk premia $\phi$ and $\phib$. The next theorem expresses this dependence through the respective drift function $G(t,s,y)$, cf.~Theorem \ref{prop-exercise1}.

\begin{theorem}\label{prop-exer-vol1} Let \begin{equation}\label{Gts_vol}G(t,s,y) := \frac{\partial  P}{\partial y} (t,s,y) \bigl(\phib(t,s,y)-\phi(t,s,y) \bigr).\end{equation} If $G(t,s,y) \leq 0$ for all $(t,s,y)$, then it is optimal not to purchase the option, i.e. $\tau^*=T$ and $L(t,s,y) =  P(t,s,y)-\byp(t,s,y)$.  If $G(t,s,y)\geq 0$ for all $(t,s,y)$, then it is optimal to purchase the option immediately, i.e.~$\tau^*=0$ and $L(t,s,y)=0$.
\end{theorem}
\begin{proof}
From (\ref{WWphi}), we observe that $dW^\phi_t = dW^\phib_t$, and $d\Wh^\Qb_t  = d\Wh^Q_t + ( \phib_t-\phi_t) \,dt.$ Therefore, the two equivalent pricing measures $Q$ and $\Qb$ are connected via the Radon-Nikodym derivative
\[Z_t :=\frac{d\Qb}{dQ} \Big|_{\F_t}  = \mathcal{E}(- \xi \Wh^\phi)_t ,\] where $\xi_t = \phib(t,S_t,Y_t) - \phi(t,S_t,Y_t)$ is the (bounded) volatility premium difference between the buyer and the market. Also, $Z$ solves the SDE: $dZ_t =- Z_{t}\,\xi_t\,d\Wh^Q_t.$
 Consequently, the process
$(Z_t \hatP_t)_{0\leq t\leq T}$ satisfies
\begin{multline}
d\, Z_t \hatP(t,S_t, Y_t)
 = \hatP(t,S_t, Y_t) \,dZ_t + Z_t\, d \hatP(t,S_t, Y_t) \\ - e^{-rt} Z_t  \hat{\rho}c(t,Y_t) (\phib(t,S_t,Y_t)-\phi(t,S_t, Y_t)) \frac{\partial  P}{\partial y}(t,S_t, Y_t)   \, dt. \label{SDE_ZPvol}
\end{multline}
Since $\hatP$, $Z$ are $Q$-martingales and $c(t,Y_t)$ is positive by convention, the process $(Z_t \hatP_t)_{0\leq t\leq T}$ is a $Q$-submartingale (resp.\ $Q$-supermartingale) if $(\phib_t-\phi_t) \frac{\partial  P}{\partial y}\geq 0$ a.s.~on $[0,T]\times \R_+ \times \R$ (resp.\ $\leq 0$). Then, it is optimal to purchase immediately (resp.\ never purchase) since the expected discounted cost $V(t,S_t, Y_t )$ increases (resp.\ decreases) over time. Finally, due to \eqref{bracket-dep}, the delayed purchase premium   admits the representation
\begin{align}L(t,s,y)= \sup_{\tau\in \setT_{t,T}} \E^\Qb\bigg\{- \int_{t}^{\tau}e^{-r(u-t)}\hat{\rho}c(u,Y_u) \underbrace{(\phib_u-\phi_u) \frac{\partial  P}{\partial y}(u,S_u,Y_u)}_{G(u,S_u, Y_u)} \,du\,|\,S_t=s, Y_t=y\bigg\}.\label{FK-fvol}
\end{align}
\end{proof}
\begin{remark}Although our analysis focuses on options written on $S$ only, Theorem \ref{prop-exer-vol1} also immediately applies for an option with payoff $F(S_T, Y_T)$. Other elements of the model, such as to unbounded risk premia,  can also be generalized as long as the martingale properties of the processes  $Z$, $\hat{P}$ and $Z \hatP$ are preserved. We do not address the full generalization here.
\end{remark}
Under the common assumption that $y \mapsto \sigma(y)$ is increasing, the drift function is again closely linked to the convexity of the option price, cf.~Corollary \ref{cor:convex-def}.
\begin{corollary}\label{cor:convexvol} Assume the option price $P(t,s,y)$ is convex in $s\in \R_+$ for every $(t,y)\in [0,T]\times\R$ and $\sigma'(y) > 0$. If $\phib(t,s,y)\leq \phi(t,s,y)$ for all $(t,s,y)$, then it is optimal to never purchase the option. If $\phib(t,s,y)\geq \phi(t,s,y)$ for all $(t,s,y)$, then it is optimal to purchase immediately.
\end{corollary}
\begin{proof}By Theorem 3.1 of \cite{RomanoTouzi97}, if the conditions of Corollary \ref{cor:convexvol} are satisfied, then the option price is increasing with respect to the volatility level, i.e.\ $\frac{\partial  P}{\partial y}(t,s,y)\geq 0$. Therefore, the corollary follows from Theorem \ref{prop-exer-vol1}.
\end{proof}
When the option payoff $F:\R_+ \mapsto \R_+$ is convex, such as the Call and the Put, then the option price is also convex (see Proposition 4.3 of \cite{RomanoTouzi97}), so Corollary \ref{cor:convexvol} applies.

By inspecting the probabilistic representation  \eqref{FK-fvol}, we deduce that if $G(t,s,y) < 0$, then the buyer should postpone her purchase since an infinitesimal reward can be obtained by waiting for an infinitesimal moment. Hence, along the exercise boundary $(s^*(t), y^*(t))$, $t\in[0,T]$, we must have $G(t,s^*(t), y^*(t)) \ge 0$. For options with convex payoffs as in Corollary \ref{cor:convexvol}, $G(t,s^*(t), y^*(t))  >0$ if and only if $\phib(t,s^*(t), y^*(t)) - \phi(t,s^*(t), y^*(t)) > 0$, so  in the exercise region the buyer must overestimate the volatility risk premium relative to the market.

\begin{corollary}\label{cor:convexvol_phi} Assume the option's payoff function $F:\R_+ \mapsto \R_+$ is convex and $\sigma'(y) > 0$. The buyer will not buy the option at $(t,s,y)$ if $\phib(t,s,y)< \phi(t,s,y)$ at that point.
\end{corollary}

Next, using Theorem \ref{prop-exer-vol1} and Corollary \ref{cor:convexvol}, we can also compare the optimal purchase strategy between vanilla options and some exotic options.

 \begin{example} (Call vs Bull Spread, cf.~Example \ref{bull-spread}) Suppose $\phib(t,s,y)\geq \phi(t,s,y)$. Since $\frac{\partial  P}{\partial y}(t,s,y)\geq 0$ for Calls, it follows that  the drift function of the Call with strike $K$ dominates that of the bull spread with strikes $(K, K_h)$, $K<K_h$. Therefore, by (\ref{tau-defn}) and  (\ref{FK-fvol}), the buyer will purchase the bull spread later than the Call. By Corollary \ref{cor:convexvol} the buyer will purchase the Call now, but may delay to buy the bull spread. Conversely, when $\phib(t,s,y) \leq \phi(t,s,y)$,  the buyer will never purchase the Call but may buy the bull spread prior to expiration.  \end{example}

\begin{example}(Price Ordering by the $q$-Optimal Measures) Intuitively, the option price should influence the buyer's purchase timing. To illustrate this, we consider the price ordering via the $q$-optimal measures studied by \cite{HHHS}. We recall that $q$-optimal measures arise from taking the probability measure $Q^{(q)}$ that minimizes the $q$-th moment of the Radon-Nikodym derivative between $Q$ and  $\Ph$, i.e.~$Q^{(q)} = \argmin_{Q} \E\{(\frac{dQ}{d\Ph})^q\}$. Pricing under $Q^{(q)}$ can also be interpreted as marginal indifference price of a risk-averse agent with a constant relative risk aversion (power) utility $U(x) = x^{q/(1-q)} \frac{1-q}{q}, \,\,q<1$.

The respective market price of volatility risk $\phi^{(q)}(t,s,y)$ is in general a complicated expression given as solution of a semi-linear PDE (see \cite{Hobson04}). However, in the case of a Heston stochastic volatility model, namely,
\begin{align*}
\left\{ \begin{aligned} dS_t &= \alpha Y_t S_t \,dt + \sqrt{Y_t} S_t \, dB_t, \\
dY_t &= 2\kappa( m - Y_t) \,dt + 2 \beta \sqrt{Y_t} \, dW_t,
\end{aligned} \right.
\end{align*}
with $\alpha, \beta, \kappa, m$ constants and $d[B,W]_t = \rho dt$, \cite{HHHS} showed that $q \mapsto \phi^{(q)}(t,s,y)$ is increasing. Therefore, assuming $Q \equiv Q^{(q_1)}$ and $\Qb \equiv Q^{(q_2)}$ with $q_1<  q_2$ (the investor is more risk averse than the market), it follows that the market Call/Put price always exceeds the investor's price and the buyer can never profit from buying from the market, so $\tau^* = T$. Conversely, if $q_1 > q_2$ then $\tau^* = 0$.
\end{example}

In general,  one has to numerically solve the free boundary problem associated with $V(t,s,y)$, namely
\begin{equation}\begin{cases}\displaystyle
\min \bigl(  \frac{\partial  V}{\partial t} + \tilde{\L}_{SY} V, P -  V \bigr) = 0, \quad 
\text{ for } (t,s,y) \in [0,T)\times \R_+\times \R, \\ 
\displaystyle V(T,s,y) = F(s), \text{  for } (s,y)\in \R_+\times \R,
\end{cases}\label{VI_Vvol}\end{equation}where we have suppressed $(t,s,y)$ and $\tilde{\L}_{SY}$ is the elliptic differential operator given by
\begin{multline*}
\tilde{\L}_{SY} f = r s \frac{\partial f}{\partial s} + \frac{1}{2}\sigma^2(y) s^2 \frac{\partial^2 f}{\partial s^2}  \\ +\left(b(t,y) - \rho c(t,y) \kappa(t,y) - \hat{\rho} c(t,y) \tilde{\phi}(t,s,y) \right) \frac{\partial f}{\partial y} + \frac{1}{2} c^2(t,y) \frac{\partial^2 f}{\partial y^2}  + \rho \sigma(y) c(t,y)
\frac{\partial^2 f}{\partial s \partial y} f - r f.
\end{multline*}
Equivalently, one can solve for the delayed purchase premium $L(t,s,y)$ via
\begin{equation}\max \bigl(    \frac{\partial L}{\partial t} + \tilde{\L}_{SY} L   - r L -\hat{\rho} c (\phib-\phi) \frac{\partial P}{\partial y}, \,-L\,\bigr) = 0, \quad 
\text{ for } (t,s,y) \in [0,T)\times \R_+\times \R,
\label{VI_L}\end{equation}with terminal condition $L(T,s,y) = 0$ for $(s,y)\in \R_+\times \R$. Compared to (\ref{VI_Vvol}), the free boundary problem (\ref{VI_L}) has a source term but a zero obstacle. Standard methods imply that under our assumptions $V$ (resp.~$L$) are the unique viscosity solutions of \eqref{VI_Vvol} and \eqref{VI_L}, and therefore the usual finite-difference methods can be applied to numerically solve (\ref{VI_Vvol}) or (\ref{VI_L}) for the associated purchase boundary that represents the critical values of $(S,Y)$ at which the option should be purchased.

We have discussed the buyer's optimal timing problem under a stochastic volatility model. In the model (\ref{SDESvol})-(\ref{SDEY}), the process $Y$ can also represent a generic non-traded stochastic factor, not necessarily for stochastic volatility. Depending on the context, one may modify the model parameters and assumptions, and therefore, the convexity of option prices may no longer play a crucial role as seen in this section. However, Theorem \ref{prop-exer-vol1} and representation \eqref{FK-fvol} still apply and can be used to infer the buyer's optimal timing.

\section{Further Applications}\label{sec:conclude}
The optimal timing problem we have discussed here also arises in other financial applications.

\subsection{Optimal Rolling for Long-Dated Options}
In a common transaction, an investor issues a long-dated option in a bespoke over-the-counter contract. This long-dated option is not traded in the market; thus, to hedge the resulting short position the investor (the hedger) instead purchases the same option with shorter maturity. For instance, to hedge  a $T=5$-year Put position on $S$, the investor might initially buy a $T_1=3$-year LEAPS Put. At a later date $\tau \le T_1$, the investor then plans to \emph{roll-over} her position into the 5-year Put, by simultaneously buying a Put expiring at $T$ and selling the Put expiring at $T_1$. In this example, assuming that LEAPS contracts are trading up to 3 years out, if the roll takes place during year 3, then a single such roll-over is needed; for maturities over 6 years multiple rolls would be required.

 Let $P_t(T)$
be the market price of a Put with arbitrary maturity $T$. Then the goal of the investor is to minimize the net cost at the roll date $\tau$ given by  $h_\tau :=P_\tau(T) - P_\tau(T_1)$, for $T_1<T$. The roll-over must take place between $\tilde{T} = T-T_1$, when the option expiring at $T$ first becomes available, and $T_1$ when the short-dated option matures.

In a complete market with a unique pricing measure $Q$, the  cost process $(h_t)_{t\geq 0}$ is a $Q$-martingale, so any admissible rolling time $\tau \in [\tilde{T}, T_1]$ will lead to the \emph{same} expected cost under $Q$. However, if the market is incomplete and the investor prices with her own pricing measure $\Qb$, then the rolling problem at time $t\leq \tilde{T}$ is
\[V^{Roll}_t = \essinf_{\tau\in \setT_{\tilde{T}, T_1}} \E^{\Qb} \left\{e^{-r(\tau-t)} h_\tau\,|\,\F_t \right\},\qquad\text{where}\qquad \setT_{\tilde{T}, T_1}  = \{ \tau \in \setT\,:\, \tilde{T}\leq \tau\leq  T_1\}.\]
The above rolling problem matches closely our purchase timing model \eqref{valfn}. Proposition \ref{prop-pc} implies that the rolling of Puts and Calls with the same strike and maturities yield identical optimal strategies. Furthermore, we can express the \emph{delayed rolling premium} as
\[ L^{Roll}_t = \esssup_{\tau\in \setT_{\tilde{T}, T_1}} \E^\Qb \left\{ - \int_{\tilde{T}}^\tau e^{-r (s-t)}(Z_s)^{-1} (d[P(T_1), Z]_s- d[P(T), Z]_s)\,\big|\,\F_t \right\}.\]
Hence, the rolling strategy depends on the difference between the drift functions $[P(T_1), Z]- [P(T), Z]$.
In both classes of models in Section \ref{sect-defaultS} and Section \ref{sect-stochV}, the resulting rolling boundary is typically nontrivial. Indeed, in contrast to Corollary \ref{cor:convex-def} in the defaultable equity model, the drift function difference $G(t,s; T)-G(t,s; T_1)$ does not take constant sign even if $\lab-\la$ is constant. In fact, the shape of $P(t,s;T) - P(t,s;T_1)$ looks like that of the short digital Call and therefore a similar non-monotone exercise boundary for $J(t,s)$ is obtained as in Figure \ref{fig:digital-default}. Similarly, in the stochastic volatility case, $\frac{\partial  P}{\partial y}(t, s,y; T) - \frac{\partial  P}{\partial y}(t,s,y;T_1)$ changes signs for different $s$ depending on the parameters.

\begin{remark} The investor's total expected discounted cost at time $0$ is $P_0(T_1) +  V^{Roll}_0$. The first part is the cost of acquiring the $T_1$-Put initially, but due to its independence of $\tau$, it is irrelevant to the selection of the optimal rolling time.  In practice, the investor may also control the expiration date and strike of the first Put, though we do not discuss this here.
\end{remark}

\subsection{Sequential Buying and Selling of Derivatives}

Another form of statistical arbitrage we may consider involves sequential buying and selling of the same derivative. Namely, the investor aims to generate profit by first buying the option at price $P(0,S_0)$ and then selling it at price $P(\tau, S_\tau) >P(0,S_0)$, making decisions based on model measure $\Qb$. Given $P(0,S_0)$, this problem is equivalent to maximizing the sale price $\E^\Qb \{ P(\tau,S_\tau) \}$, i.e.~\eqref{def-V} up to a sign-change. If the purchase date $\nu$ can also be optimally timed, the investor then has a two-stage timing problem,
\begin{align}\label{eq:U-problem}
U_t = \esssup_{\nu\in \setT_{t,T}, \tau\in \setT_{\nu,T}} \E^{\Qb} \left\{ e^{-r(\tau-t)} P_\tau - e^{-r(\nu-t)}P_\nu \,|\,\F_t \right\}.
\end{align}

While \eqref{eq:U-problem} is  a two-stopping-time problem, it can be straightforwardly decomposed into sequential stopping. Indeed, defining (cf.~\eqref{valfn})
\begin{align*}R_u = \esssup_{\tau\in \setT_{u,T}} \E^{\Qb} \left\{ e^{-r(\tau-u)} P_\tau \,|\,\F_u \right\}, \qquad u\in [0,T],
\end{align*}
we have for any $t\in [0,T]$
\begin{align}\label{U_R}
U_t = \esssup_{\nu\in \setT_{t,T}} \E^{\Qb} \left\{ e^{-r(\nu-t)}( R_\nu -P_\nu) \,|\,\F_t \right\}.
\end{align}
Hence, we can first numerically solve the standard optimal stopping problem for $R$ and then another one for $U$.

\begin{proposition}\label{prop-buysell}
The value function $U$ from \eqref{eq:U-problem} admits the delayed purchase premium representation
\begin{align}
U_t = \esssup_{\nu\in \setT_{t,T}, \tau\in \setT_{\nu,T}} \E^{\Qb} \left\{ \int_{\nu}^{\tau}  Z_u^{-1} e^{-r(u-t)} \, d [P, Z]_u\,|\,\F_t \right\}.
\end{align}
\end{proposition}
Proposition \ref{prop-buysell} implies that in Markovian models, the drift function $G$ is once again useful in analyzing the optimal
purchase and liquidation decisions. For instance, in the defaultable equity model of Section \ref{sect-defaultS} with the drift function $G(t,s)$ defined in \eqref{driftterm}, we have $U(t,s) = \sup_{\nu\in \setT_{t,T}, \tau\in \setT_{\nu,T}} \E^{\Qb} \left\{ \int_{\nu}^{\tau}e^{-r(u-t)} G(u,S_u)\, du |S_t=s\right\}$. Therefore, if $G$ is of constant sign, then following the spirit of Theorem \ref{prop-exercise1} either the option is never purchased (whenever $G \leq 0$) or it is purchased immediately and held till maturity (whenever $G \geq 0$). Similar conclusions can be made for the stochastic volatility setup in Section \ref{sect-stochV}.

In other cases, the investor will buy and then sell the option during $[t,T]$ and the timing strategy involves both a buy region and a subsequent sell region. Numerical solutions in a parametric model can be straightforwardly obtained using the sequential representation of $U$ in \eqref{U_R}. Finally, one can also consider more complex models of contract accumulation/liquidation following the methods of \cite{Henderson2008}.

\subsection{Other Extensions}
The optimal timing problem can also be extended in a number of directions, such as when (i) the underlying $S$ admits other dynamics, e.g.~jump diffusion; (ii) the buyer wants to purchase other financial instruments, e.g.~foreign exchange, fixed income, or credit derivatives; and (iii) the option buyer observes ask prices from \emph{multiple} sellers. In the last case, each seller prices the option under a different EMM $Q^i$, yielding a no-arbitrage price $P^i_t$. To the buyer, the cost of the option is now the cheapest offer among the seller's prices $\min_i\{P_t^i\}$, which can be viewed as the no-arbitrage price for the option under a certain EMM $\hat{Q}$, i.e., $\exists\hat{Q} $ such that $\E^{\hat{Q}}\{ e^{-r(T-t)}   F(S_T)|\F_t\} = \min_i\{P_t^i\}$. Hence, we can reduce this multiple-seller problem to the single-seller case discussed in this paper.

Finally, another practical extension is to incorporate the buyer's risk preferences in her timing problem. One common approach is to formulate the buyer's problem in terms of utility maximization. The buyer's valuation of the option can be derived from the concept of utility-indifference price (or certainty equivalent), and her purchase decision naturally depends on the dynamics of both the buyer's utility-indifference price and the market price. In a similar spirit, the recent works by \cite{aytacsircar} and \cite{leungsircar2} apply indifference pricing to study static-dynamic hedging that also involves purchasing derivatives from the market.

In the model presented, the buyer knows precisely the market pricing measure $Q$. In some situations, such as when  options are not liquidly traded, there may be in fact ambiguity about how the market/sellers generate ask quotes. Consequently, it may be useful to introduce model uncertainty regarding $Q$. These extensions will be considered in future works.

\subsection*{Acknowledgements}
We are grateful to Paul Glasserman, Matheus Grasselli and Vicky Henderson for useful conversations and suggestions, and Adi Dror for his assistance with the numerical implementation.

\bibliographystyle{jas99}

\begin{small}
\bibliography{mybib_june2010,timingOption}

\begin{thebibliography}{}

\bibitem[\protect\citeauthoryear{Allen \& Padovani}{Allen and
  Padovani}{2002}]{AllenPadovani02}
Allen, S. and O.~Padovani, 2002:
\newblock Risk management using quasi–static hedging.
\newblock {\em Economic Notes}, {\bf 31 (2)}, 277--336.

\bibitem[\protect\citeauthoryear{Alvarez \& Stenbacka}{Alvarez and
  Stenbacka}{2004}]{AlvarezStenbacka04}
Alvarez, L. and R.~Stenbacka, 2004:
\newblock Optimal risk adoption: A real options approach.
\newblock {\em Economic Theory}, {\bf 23}, 123--148.

\bibitem[\protect\citeauthoryear{Bellamy \& Jeanblanc}{Bellamy and
  Jeanblanc}{2000}]{BellamyJeanblanc00}
Bellamy, N. and M.~Jeanblanc, 2000:
\newblock Incompleteness of markets driven by a mixed diffusion.
\newblock {\em Finance and Stochastics}, {\bf 4}, 209--222.

\bibitem[\protect\citeauthoryear{Carr, Jarrow \& Myneni}{Carr
  et~al.}{1992}]{carr_jarrow_myneni92}
Carr, P., R.~Jarrow, and R.~Myneni, 1992:
\newblock Alternative characterizations of {A}merican put options.
\newblock {\em Mathematical Finance}, {\bf 2}(2), 87--106.

\bibitem[\protect\citeauthoryear{Delbaen}{Delbaen}{2006}]{Delbaen-mstable06}
Delbaen, F., 2006:
\newblock The structure of $m-$stable sets and in particular of the set of risk
  neutral measures.
\newblock In {\em Lecture Notes in Mathematics}, volume 1874. Springer Berlin /
  Heidelberg, 215-258.

\bibitem[\protect\citeauthoryear{{El Karoui}, Jeanblanc \& Jiao}{{El Karoui}
  et~al.}{2010}]{ElKaroui_J_J_aftdefault10}
{El Karoui}, N., M.~Jeanblanc, and Y.~Jiao, 2010:
\newblock What happens after a default: the conditional density approach.
\newblock {\em Stochastic processes and their applications}, {\bf 120},
  1011--1032.

\bibitem[\protect\citeauthoryear{{El Karoui} \& Quenez}{{El Karoui} and
  Quenez}{1995}]{ElKaroui1995}
{El Karoui}, N. and M.~Quenez, 1995:
\newblock Dynamic programming and pricing of contingent claims in an incomplete
  market.
\newblock {\em SIAM Journal on Control and Optimization}, {\bf 33}, 29--66.

\bibitem[\protect\citeauthoryear{F\"{o}llmer \& Schweizer}{F\"{o}llmer and
  Schweizer}{1990}]{FollmerSchweizer1990}
F\"{o}llmer, H. and M.~Schweizer, 1990:
\newblock Hedging of contingent claims under incomplete information.
\newblock In {\em Applied Stochastic Analysis, Stochastics Monographs}, Davis,
  M. and Elliot, R., editors, volume~5. Gordon and Breach, London/New York, 389
  - 414.

\bibitem[\protect\citeauthoryear{Fouque, Papanicolaou \& Sircar}{Fouque
  et~al.}{2000}]{Sircarbook}
Fouque, J.-P., G.~Papanicolaou, and R.~Sircar, 2000:
\newblock {\em Derivatives in Financial Markets with Stochastic Volatility}.
\newblock Cambridge University Press.

\bibitem[\protect\citeauthoryear{Fritelli}{Fritelli}{2000}]{Frittelli00}
Fritelli, M., 2000:
\newblock The minimal entropy martingale measure and the valuation problem in
  incomplete markets.
\newblock {\em Math. Finance}, {\bf 10}, 39--52.

\bibitem[\protect\citeauthoryear{Fujiwara \& Miyahara}{Fujiwara and
  Miyahara}{2003}]{FujiwaraMiyahara03}
Fujiwara, T. and Y.~Miyahara, 2003:
\newblock The minimal entropy martingale measures for geometric {L}\'evy
  processes.
\newblock {\em Finance and Stochastics}, {\bf 7}(4), 509--531.

\bibitem[\protect\citeauthoryear{Guo, Jarrow \& Zeng}{Guo
  et~al.}{2009}]{GuoJarrow_incomp_info_2009}
Guo, X., R.~Jarrow, and Y.~Zeng, 2009:
\newblock Credit risk models with incomplete information.
\newblock {\em Mathematics of Operations Research}, {\bf 34}(2), 320--332.

\bibitem[\protect\citeauthoryear{Henderson \& Hobson}{Henderson and
  Hobson}{2003}]{HenderHobs_JumpCompare}
Henderson, V. and D.~Hobson, 2003:
\newblock Coupling and option price comparisons in a jump-diffusion model.
\newblock {\em Stochastics and Stochastics Reports}, {\bf 75}, 79--101.

\bibitem[\protect\citeauthoryear{Henderson \& Hobson}{Henderson and
  Hobson}{2008}]{Henderson2008}
Henderson, V. and D.~Hobson, 2008:
\newblock Optimal liquidation of derivative portfolios.
\newblock {\em Math. Finance}.
\newblock To appear.

\bibitem[\protect\citeauthoryear{Henderson, Hobson, Howison \&
  T.Kluge}{Henderson et~al.}{2005}]{HHHS}
Henderson, V., D.~Hobson, S.~Howison, and T.Kluge, 2005:
\newblock A comparison of $q$-optimal option prices in a stochastic volatility
  model with correlation.
\newblock {\em Review of Derivatives Research}, {\bf 8}, 5--25.

\bibitem[\protect\citeauthoryear{Hobson}{Hobson}{2004}]{Hobson04}
Hobson, D., 2004:
\newblock Stochastic volatility models, correlation, and the {$q$}-optimal
  measure.
\newblock {\em Math. Finance}, {\bf 14}(4), 537--556.

\bibitem[\protect\citeauthoryear{\.{I}lhan \& Sircar}{\.{I}lhan and
  Sircar}{2005}]{aytacsircar}
\.{I}lhan, A. and R.~Sircar, 2005:
\newblock Optimal static-dynamic hedges for barrier options.
\newblock {\em Math. Finance}, {\bf 16}, 359--385.

\bibitem[\protect\citeauthoryear{Jarrow, Lando \& Yu}{Jarrow
  et~al.}{2005}]{JarrowLandoYu05}
Jarrow, R.~A., D.~Lando, and F.~Yu, 2005:
\newblock Default risk and diversification: theory and empirical implications.
\newblock {\em Math. Finance}, {\bf 15}(1), 1--26.

\bibitem[\protect\citeauthoryear{Karatzas \& Shreve}{Karatzas and
  Shreve}{1998}]{KaratzasShreve01}
Karatzas, I. and S.~Shreve, 1998:
\newblock {\em Methods of Mathematical Finance}.
\newblock Springer.

\bibitem[\protect\citeauthoryear{Kramkov}{Kramkov}{1996}]{Kramkov_decompose}
Kramkov, D., 1996:
\newblock Optional decomposition of supermartingales and hedging contingent
  claims in incomplete security markets.
\newblock {\em Probability Theory and Related Fields}, {\bf 105}, 459--479.

\bibitem[\protect\citeauthoryear{Leung \& Sircar}{Leung and
  Sircar}{2009}]{leungsircar2}
Leung, T. and R.~Sircar, 2009:
\newblock Exponential hedging with optimal stopping and application to {ESO}
  valuation.
\newblock {\em SIAM Journal of Control and Optimization}, {\bf 48}(3),
  1422--1451.

\bibitem[\protect\citeauthoryear{Linetsky}{Linetsky}{2006}]{Linetsky06}
Linetsky, V., 2006:
\newblock Pricing equity derivatives subject to bankruptcy.
\newblock {\em Math. Finance}, {\bf 16}(2), 255--282.

\bibitem[\protect\citeauthoryear{Merton}{Merton}{1976}]{Merton_jump1976}
Merton, R., 1976:
\newblock Option pricing when underlying stock returns are discontinuous.
\newblock {\em Journal of Financial Economics}, {\bf 3}, 125--144.

\bibitem[\protect\citeauthoryear{Oksendal \& Sulem}{Oksendal and
  Sulem}{2005}]{OksendalSulemBook}
Oksendal, B. and A.~Sulem, 2005:
\newblock {\em Applied Stochastic Control of Jump Diffusions}.
\newblock Springer.

\bibitem[\protect\citeauthoryear{Peskir, Glover \& Samee}{Peskir
  et~al.}{2009}]{PeskirSameeAsian}
Peskir, G., K.~Glover, and F.~Samee, 2009:
\newblock The {B}ritish {A}sian option.
\newblock {\em Sequential Anal.} to appear.

\bibitem[\protect\citeauthoryear{Peskir \& Samee}{Peskir and
  Samee}{2008}]{PeskirSameePut}
Peskir, G. and F.~Samee, 2008:
\newblock The {B}ritish put option.
\newblock Technical Report Research Report No. 1, Probab. Statist. Group
  Manchester.

\bibitem[\protect\citeauthoryear{Peskir \& Shiryaev}{Peskir and
  Shiryaev}{2006}]{Peskir-Shiryaev-book}
Peskir, G. and A.~N. Shiryaev, 2006:
\newblock {\em Optimal Stopping and Free-boundary problems}.
\newblock Birkhauser-Verlag, Lectures in Mathematics, ETH Zurich.

\bibitem[\protect\citeauthoryear{Pham}{Pham}{1997}]{PhamAmer}
Pham, H., 1997:
\newblock Optimal stopping, free boundary, and {A}merican option in a
  jump-diffusion model.
\newblock {\em Applied Mathematical and Optimization}, {\bf 35}, 145--164.

\bibitem[\protect\citeauthoryear{Protter}{Protter}{2003}]{ProtterBook}
Protter, P., 2003:
\newblock {\em Stochastic integration and differential equations}.
\newblock Springer.

\bibitem[\protect\citeauthoryear{Riedel}{Riedel}{2009}]{Riedel_stopAmbig09}
Riedel, F., 2009:
\newblock Optimal stopping with multiple priors.
\newblock {\em Econometrica}, {\bf 77}(3), 857 -- 908.

\bibitem[\protect\citeauthoryear{Romano \& Touzi}{Romano and
  Touzi}{1997}]{RomanoTouzi97}
Romano, M. and N.~Touzi, 1997:
\newblock Contingent claims and market completeness in a stochastic volatility
  model.
\newblock {\em Math. Finance}, {\bf 7}(4), 399--410.

\bibitem[\protect\citeauthoryear{Wilmott, Howison \& Dewynne}{Wilmott
  et~al.}{1995}]{WillmottHowisonDewynne95}
Wilmott, P., S.~Howison, and J.~Dewynne, 1995:
\newblock {\em The Mathematics of Financial Derivatives}.
\newblock Cambridge University Press.

\end{thebibliography}
\end{small}

\end{document}